\documentclass[nature,%
,showpacs
,superscriptaddress
]
{revtex4-2}

\usepackage{etoolbox}

\usepackage{graphicx,subfigure,bbm}

\usepackage{amsmath}
\usepackage{tabularx}
\usepackage{algorithm}
\usepackage{algpseudocode}
\usepackage{float}
\usepackage{nicefrac}        
\usepackage{amsthm}
\usepackage{amssymb}
\usepackage{array,bm}
\usepackage{graphicx}
\usepackage{fancyhdr}
\usepackage{booktabs,threeparttable}
\usepackage{siunitx}
\usepackage{color}
\usepackage{extarrows}
\usepackage{enumerate}
\usepackage{epstopdf}
\usepackage[colorlinks,
            linkcolor=black,
            citecolor=black,
            urlcolor=blue
            ]{hyperref}
\usepackage[latin1]{inputenc}
\usepackage{tikz}
\usetikzlibrary{shapes,arrows}

\tikzstyle{decision} = [diamond, draw, fill=blue!20,
    text width=4.5em, text badly centered, node distance=3cm, inner sep=0pt]
\tikzstyle{block} = [rectangle, draw, fill=blue!20,
    text width=10em, text centered, rounded corners, minimum height=4em]
\tikzstyle{line} = [draw, -latex']
\tikzstyle{cloud} = [rectangle, draw,fill=red!20, node distance=7cm,
    minimum height=4em]

\newcommand{\be}{\begin{equation}}
\newcommand{\ee}{\end{equation}}
\newcommand{\bea}{\begin{eqnarray}}
\newcommand{\eea}{\end{eqnarray}}

\DeclareMathAlphabet{\varmathbb}{U}{bbold}{m}{n}

\definecolor{C0}{HTML}{1f77b4}
\definecolor{C1}{HTML}{ff7f0e}
\definecolor{C2}{HTML}{2ca02c}
\definecolor{C3}{HTML}{d62728}
\definecolor{C4}{HTML}{9467bd}
\definecolor{C5}{HTML}{8c564b}

\begin{document}
\title{Scalable Physics-Inspired Transformers for Spin Glasses}

\author{Lu Zhong}
\altaffiliation{These authors contributed equally}
\affiliation{Institute of Fundamental and Frontier Sciences, University of Electronic Science and Technology of China, Chengdu 611731, China}
\author{Wenli Duan}
\altaffiliation{These authors contributed equally}
\affiliation{Institute of Fundamental and Frontier Sciences, University of Electronic Science and Technology of China, Chengdu 611731, China}

\author{Jing Liu}
\email[Corresponding authors: ]{jing.liu@bupt.edu.cn}
\affiliation{School of Physical Science and Technology, Beijing University of Posts and Telecommunications, Beijing 100876, China}
\affiliation{Institute of Theoretical Physics, Chinese Academy of Sciences, Beijing 100190, China}

\author{Pan Zhang}
\email[Corresponding authors: ]{panzhang@itp.ac.cn}
\affiliation{Institute of Theoretical Physics, Chinese Academy of Sciences, Beijing 100190, China}
\affiliation{School of Fundamental Physics and Mathematical Sciences, Hangzhou Institute for Advanced Study, UCAS, Hangzhou 310024, China}

\author{Ying Tang}
\email[Corresponding authors: ]{jamestang23@gmail.com}
\affiliation{Institute of Fundamental and Frontier Sciences, University of Electronic Science and Technology of China, Chengdu 611731, China}
\affiliation{School of Physics, University of Electronic Science and Technology of China, Chengdu 611731, China}
\affiliation{Key Laboratory of Quantum Physics and Photonic Quantum Information, Ministry of Education, University of Electronic Science and Technology of China, Chengdu 611731, China}
\affiliation{Non-classical Information Science Basic Discipline Research Center of Sichuan Province, University of Electronic Science and Technology of China, Chengdu 611731, China}

\begin{abstract}
Efficient sampling of the Boltzmann distribution in frustrated spin glasses is central to statistical mechanics and combinatorial optimization. Despite advances in machine-learning-based approaches, two issues persist: limited gains from increased model scale; and high computational cost on large systems that limits practical efficiency. Here, we develop a physics-inspired transformer with interaction-aware attention and concatenated positional embeddings, coupled to population-assisted training to address these challenges. By further leveraging FlashAttention and key-value caching for efficient autoregressive sampling, it achieves up to $36.4\times$ speedup over vanilla variational autoregressive networks, enabling neural-network simulations of spin-glass systems with up to $4096$ spins on a single GPU.
Across the Sherrington-Kirkpatrick and two- and three-dimensional Edwards-Anderson models, FlashVAN can resolve full probability distributions and free energies, and generate nontrivial finite-temperature overlap statistics in agreement with independent numerical benchmarks. For ground-state searches, FlashVAN also exhibits instance-dependent advantages complementary to those of Global Annealing (GA), a state-of-the-art machine-learning-assisted optimization approach, while our training strategy can achieve a $3\times$ speedup over GA on a particularly challenging 3D instance. This framework thus establishes a scalable paradigm for frustrated spin-glass systems.
\end{abstract}

\maketitle

\section{Introduction}

Understanding the collective behavior of high-dimensional spin-glass models remains a central problem in statistical mechanics and combinatorial optimization~\cite{RevModPhys.58.801,PhysRevLett.52.1156,schuetz2022combinatorial,sherrington202550}. Paradigmatic spin-glass models, such as the Sherrington-Kirkpatrick (SK)~\cite{PhysRevLett.35.1792} and Edwards-Anderson (EA)~\cite{edwards1975theory}, exhibit  frustration, disorder, and rugged energy landscapes. Characterizing these systems requires sampling probability distributions over an exponentially large configuration space. At finite temperature, physical observables depend on the relative statistical weights of competing configurations. Conventional sampling methods, including Markov chain Monte Carlo~\cite{binder1992monte}, can face severe equilibration challenges in rugged energy landscapes, particularly at low temperatures where relaxation times become extremely long. Consequently, understanding the free-energy landscape, ground state, and phase structure of general spin glasses remains an outstanding theoretical and computational problem in physics. 

Recent advances in machine learning have opened new possibilities for modeling complex spin-glass systems. In particular, variational autoregressive networks (VANs) \cite{PhysRevLett.122.080602} offer a tractable representation of the Boltzmann distribution, enabling configuration-probability evaluation and direct autoregressive sampling. VANs and related generative models have been applied to equilibrium statistical mechanics \cite{mehta2019high,PhysRevE.101.053312}, quantum many-body systems \cite{carleo2017solving,RevModPhys.91.045002,PhysRevResearch.2.023358,TomWesterhout2020GeneralizationPO,PhysRevLett.128.090501,NetKet, science9774}, and nonequilibrium dynamics \cite{tang2023neural,tang_learning_2024,weng2025tracking}. Yet a critical limitation lies in model scalability: increasing network size or depth does not automatically improve the representation of spin-glass distributions. Expressive architectures must also incorporate appropriate physical structure and remain computationally feasible to train and sample. Architectures such as the Masked Autoencoder for Distribution Estimation (MADE)~\cite{germain2015made} and the neural autoregressive distribution estimator (NADE)~\cite{LarochelleNade,NADE} have therefore been widely used, but representing larger systems and complex low-temperature distributions remains challenging. In addition, training based solely on model-generated samples can reinforce sampling biases when relevant configurations are insufficiently explored. 

Besides modeling the full probability distribution at finite temperature, the scalability bottleneck also limits machine-learning methods in the task of searching for ground states at low temperature. Generative models have also been combined with Monte Carlo methods, with their effectiveness depending on the sampling problem~\cite{ciarella2023machine}. Approaches exploiting nearest-neighbor structure~\cite{del2025nearest} and Global Annealing (GA)~\cite{del2026demonstrating}, which uses model-generated configurations within a Monte Carlo optimization procedure, illustrate complementary developments. In parallel, reinforcement-learning approaches for searching spin-glass ground states at low temperatures have also been proposed \cite{fan2023searching}, although their accuracy was debated \cite{boettcher2023deep,fan2023reply}. These developments raise an important question: can an autoregressive generative framework achieve competitive ground-state optimization and accurate finite-temperature distribution learning at a practical computational cost? Establishing such a framework requires both efficient modeling and validation against established optimization and equilibrium-sampling methods.

Here, we introduce FlashVAN, a Transformer-based autoregressive framework combining physics-inspired attention, concatenated positional embeddings, population-assisted training, and hardware-efficient implementation. The attention design exploits the boundary-dependent conditional structure of lattice models, also used in TwoBo~\cite{Biazzo_Wu_Carleo_2024}, to guide sparse causal attention and further accelerate training for spatially structured systems, including the 2D and 3D EA models. The fully connected SK model retains full causal attention. We also introduce a concatenated positional embedding that preserves spin-state and positional information in separate feature dimensions, improving learning stability. To strengthen learning in challenging low-temperature regimes, we develop population-assisted training: a persistent sample pool couples model-generated configurations with standard local Metropolis Monte Carlo (LMC) refinement~\cite{Metropolis1953}, supplying learning signals acquired through Monte Carlo exploration. The refined population is subsequently distilled into the autoregressive model, retaining direct sampling after training. While GA primarily employs a generative model to assist Monte Carlo optimization, our population-assisted strategy uses Monte Carlo refinement to train a directly sampleable autoregressive distribution. The resulting model supports both low-energy search and repeated finite-temperature measurements after training. We further include a hardware-efficient implementation: by leveraging FlashAttention kernels together with key-value (KV) caching for efficient autoregressive sampling, as developed for modern transformers~\cite{vaswani2017attention,dao2022flashattention,shah2024flashattention}, the present framework makes training and sampling practical at substantially larger scales. FlashVAN thus addresses the complementary challenges of probability representation, configuration-space exploration, and computational efficiency.

Across the SK and two- and three-dimensional EA models, FlashVAN demonstrates competitive ground-state optimization and reproduces nontrivial finite-temperature overlap statistics (Figs.~\ref{fig:2}-\ref{fig:4}). Independent comparisons with established optimization and equilibrium-sampling methods assess its energy accuracy and overlap structure, including a matched GA benchmark with training costs included. The optimization comparison reveals complementary instance-dependent advantages, while applying our training strategy to a simpler MADE backbone further accelerates a challenging instance. Training and evaluation of a $4096$-spin generative sampler on a single GPU further demonstrate the scale accessible within this hardware budget (Table~\ref{tab:performance}). Together, these results connect advances in neural autoregressive modeling to the study of low-temperature spin-glass statistics.

\begin{figure}[ht!]
	\begin{center}
		\includegraphics[width=\columnwidth]{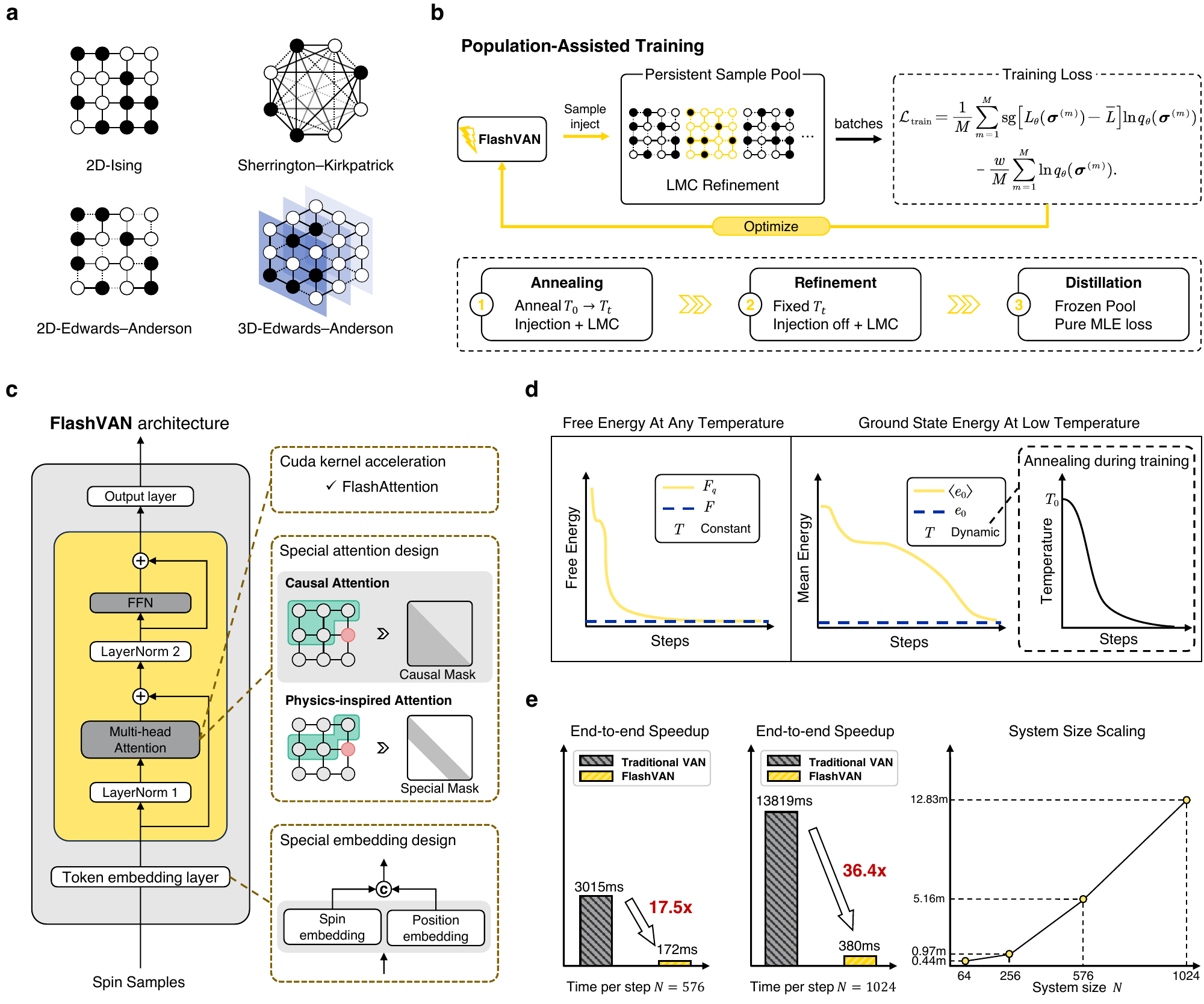}
	\end{center}
	\caption{\textbf{The present framework facilitates solving spin-glass problems.}
(a) Schematic illustration of representative spin-glass systems.
(b) Population-assisted training: annealing, fixed-temperature refinement, and final likelihood-based distillation. Local Metropolis Monte Carlo (LMC) updates refine the sample pool, while the training loss used during annealing and refinement is defined in Eq.~\ref{eq:Loss}.
(c) Architecture: featuring CUDA kernel acceleration modules, a physics-inspired attention mechanism, and a tailored positional embedding design, which together enhance expressivity and accuracy.
(d) Training results show the convergence of the free energy and ground-state energy under the temperature-annealing scheme.
(e) Computational performance for the 2D EA model, demonstrating a substantial speedup over vanilla VAN approaches including faster sampling procedures, and favorable scaling of computational time with system sizes.}
	\label{fig:1}
\end{figure}

\begin{table}[ht!]
    \centering
    \begin{threeparttable}
        \caption{Comparison of the largest system size reported for the machine-learning methods on the benchmark models. Approaches include the vanilla VAN~\cite{PhysRevLett.122.080602}, variational neural annealing (VNA)~\cite{VNA_2021}, ng-VAN~\cite{PhysRevE.111.025304}, two-body interactions (TwoBo)~\cite{Biazzo_Wu_Carleo_2024}, nearest-neighbors neural network (4N)~\cite{del2025nearest}, GA~\cite{del2026demonstrating}, and our FlashVAN (on a single GPU). Values represent the maximum system size $N$ reported. NA: Not available, as it was not reported in the original literature.}
        \label{tab:performance}
        \small
        \begin{tabular*}{\textwidth}{@{\extracolsep{\fill}}lccccccc}
            \toprule
            Model & VAN & VNA & ng-VAN & TwoBo & 4N & GA & FlashVAN \\ 
            \midrule
            SK     & 100 & 100 & 30  & NA   & NA  & NA   & 256  \\
            2D EA  & NA  & 3,600 & NA  & 1,024 & 256 & NA   & 3,600 \\
            3D EA  & NA  & NA   & NA  & 1,728 & NA  & 2,744 & 4,096 \\
            \bottomrule
        \end{tabular*}
    \end{threeparttable}
\end{table}

\section{Results}

\subsection{Framework}
Consider a spin system of size $N$, whose configurations $\boldsymbol{\sigma}\in\{\pm1\}^N$ follow the Boltzmann distribution
\begin{equation}
    p(\boldsymbol{\sigma}) = \frac{e^{-\beta E(\boldsymbol{\sigma})}}{Z},
\end{equation}
where $\beta = 1/T$  is the inverse temperature and $Z$ is the partition function. The VAN~\cite{PhysRevLett.122.080602}  can approximate this target distribution through a tractable autoregressive factorization, $q_\theta(\boldsymbol{\sigma}) = \prod_{i=1}^{N} q_\theta(\sigma_i \,|\, \sigma_{<i})$, allowing for direct sampling. By minimizing the Kullback-Leibler (KL) divergence between $q_\theta$ and $p$, VAN connects probability modeling to variational estimation of the free energy $F = -(1/\beta)\ln Z$ and, through low-temperature annealing, to the search for ground-state configurations with energy $E_0 = \min_{\boldsymbol{\sigma}} E(\boldsymbol{\sigma})$.

To address the challenges of representing complex spin distributions and scaling autoregressive models to larger systems, we develop the Flash Variational Autoregressive Network (FlashVAN). The framework combines physics-inspired attention mechanisms, tailored positional representations, and a population-assisted training strategy. This strategy couples model-generated samples with local Metropolis Monte Carlo updates~\cite{Metropolis1953} through a compact sample pool, combining autoregressive generation with Monte Carlo refinement of a persistent population. Hardware-efficient implementations based on FlashAttention kernels~\cite{dao2022flashattention,shah2024flashattention} and key-value (KV) caching further accelerate model evaluation and autoregressive generation. Figure~\ref{fig:1} summarizes the framework, with its components detailed below.

\subsubsection{Physics-inspired sparse attention}

The design of our attention mechanism is physically rooted in the locality of interactions in lattice spin models. Standard autoregressive Transformer architectures employ full causal self-attention, which evaluates  dependencies across the entire generated sequence. For nearest-neighbor models, however, the conditional probability of the next spin depends on the boundary separating the sampled and unsampled regions. This boundary-dependent structure has been exploited in previous approaches, including TwoBo~\cite{Biazzo_Wu_Carleo_2024} and tensor network Monte Carlo algorithms~\cite{frias-perez2023collective,chen2025tnmc,chen2025batchtnmc}. TwoBo represents this structure through coupling-weighted effective fields and a feedforward network, whereas FlashVAN uses it to guide the design of sparse causal attention in a Transformer.

More specifically, for a 2D nearest-neighbor model with open boundary conditions and raster-scan ordering, the relevant boundary contains $\mathcal{O}(L)$ spins; its counterpart in 3D scales as $\mathcal{O}(L^2)$. Guided by this structure, FlashVAN restricts causal attention to a local window containing the relevant boundary spins (Fig.~\ref{fig:1}c), reducing computational cost while retaining the required conditioning information. The ordering, boundary conditions, and tokenization determine the appropriate attention mask, as detailed in Methods and Supplementary Section~I.A. For the fully connected SK model, where interactions extend across all pairs of spins, this lattice-based truncation does not apply. FlashVAN therefore employs full causal attention.

\subsubsection{\texorpdfstring{Concatenated positional embedding}{Concatenated positional embedding}}

We find that the performance of VANs is highly sensitive to the representation of positional information. To investigate this sensitivity, we systematically evaluated a range of positional encoding strategies across physical models with varying topological complexities. For the Ising model, which is characterized by a simple lattice structure, a learnable additive (absolute) positional embedding is sufficient. However, for the more intricate Sherrington-Kirkpatrick (SK) model, additive embeddings induce pronounced oscillations in the variational free-energy convergence at low temperatures (Supplementary Fig.~4). One possible explanation is that, for networks with limited depth, encoding positional variance solely through additive offsets tends to conflate spatial and state information, undermining the model's ability to resolve individual spins and triggering these instabilities. To overcome this, we design a concatenated positional embedding (Fig.~\ref{fig:1}c) that preserves spin and positional information in separate feature dimensions, rather than combining them through addition. This approach improves convergence stability and accuracy in the tested comparisons.

These observations highlight the importance of how positional information is incorporated into transformer-based VANs.  For more structured spin systems, such as the 2D and 3D Edwards-Anderson (EA) models, incorporating relative positional information yields additional empirical gains, suggesting that relative positional embeddings, such as 2D or 3D rotary positional embeddings (RoPE)~\cite{RoPE}, are also a promising design choice. Here, RoPE complements the concatenated input representation by supplying relative positional information within attention. A more detailed description of the positional representations and tokenization is provided in Methods Sec.~\ref{sec:concatePE}.

\subsubsection{Overall architecture}

Figure~\ref{fig:1}c illustrates the overall architecture of FlashVAN, which integrates a tailored concatenated positional embedding and physics-inspired sparse attention into a Transformer-based autoregressive model. As in other Transformer-based neural ansatzes~\cite{rende2024simple,sprague2024variational,van2025many}, the model consists of a stack of decoder layers, each containing masked multi-head self-attention and a feed-forward network (FFN). Spin configurations are represented through the concatenated positional embedding and processed by the decoder layers, where the attention pattern is determined by the interaction structure of the physical system. The output layer produces normalized conditional probabilities, enabling direct autoregressive sampling. The attention computation is detailed in Supplementary Section~I.A.

To accelerate computation, we employ FlashAttention-2 CUDA kernels~\cite{Dao_2024} to reduce memory bandwidth overhead during attention computation. In addition, we integrate a key-value (KV) caching strategy to improve autoregressive sampling efficiency. Implementation details are provided in Methods Section~\ref{sec:hardware_acceleration}.

\subsubsection{\texorpdfstring{Population-assisted training}{Population-assisted training}}

Learning the Boltzmann distribution of a spin glass presents a challenge beyond identifying low-energy configurations: finite-temperature observables depend on the relative statistical weights of competing free-energy valleys. Conventional VAN training minimizes the variational free energy using model-generated samples, but this self-sampling procedure can reinforce sampling biases when important regions of configuration space are insufficiently explored. To address this limitation, we develop a population-assisted training strategy that couples FlashVAN with a persistent Monte Carlo-refined sample pool (Fig.~\ref{fig:1}b).

The central idea is to establish a feedback loop between autoregressive generation and Monte Carlo refinement. Model-generated configurations replenish the pool, local Metropolis Monte Carlo updates refine its distribution, and the updated population provides additional learning signals to FlashVAN. Generative models have also been combined with Monte Carlo methods in previous work, such as Global Annealing (GA)~\cite{del2026demonstrating}, where model-generated configurations serve as global proposals within a Monte Carlo optimization procedure. Whereas GA uses the model to assist Monte Carlo, FlashVAN uses Monte Carlo to assist variational autoregressive learning, with the aim of obtaining a directly sampleable generative model that can be reused for physical measurements beyond low-energy search.

After initialization on a high-temperature population and selection of a smaller persistent pool, we organize this strategy into three stages: \textbf{annealing, refinement, and distillation}. During annealing, the model and sample pool co-evolve as the temperature decreases. At the target temperature, refinement suspends model-sample injection and population resampling, allowing further local relaxation while model training continues. In the final distillation stage, the refined pool is frozen and its statistical information is transferred into the same autoregressive model through maximum-likelihood training. Once trained, FlashVAN generates independent samples from its learned distribution without additional Monte Carlo updates.

We apply this training strategy to the SK, 2D EA, and 3D EA models, with task-dependent configurations for finite-temperature observables and ground-state search. In particular, we demonstrate its effectiveness in three-dimensional spin glasses by reproducing finite-temperature equilibrium overlap statistics across the investigated temperature range, as validated against established Monte Carlo benchmarks below. The training objectives and update rules are detailed in Methods Sec.~\ref{sec:TrainingStrategy}. Supplementary Section~II.B describes initialization and the three stages; task-specific settings are listed in Supplementary Tables~2, 4, and~7.

\subsection{Application to spin glass systems}

We evaluate FlashVAN on three representative spin-glass systems: the Sherrington-Kirkpatrick (SK) model and the Edwards-Anderson (EA) models in two and three dimensions. Table~\ref{tab:performance} compares the largest system sizes reported for representative machine-learning approaches, highlighting the scalability of FlashVAN within this class of methods. Beyond computational scalability, we assess its performance through complementary tests of energy optimization and finite-temperature equilibrium sampling (Figs.~\ref{fig:2}-\ref{fig:4}). These evaluations include comparisons with established non-neural-network methods, providing complementary benchmarks for optimization accuracy and equilibrium sampling. Energy comparisons evaluate the accuracy of low-energy solutions, while overlap distributions probe whether the learned sampler captures the statistical weights of competing configurations.

For two independent replicas drawn at the same temperature with identical couplings $J$, the spin overlap is defined as
\begin{equation}\label{eq:overlap_distribution}
    q = \frac{1}{N}\sum_{i=1}^{N}\sigma_i^1\sigma_i^2.
\end{equation}
We denote its equilibrium distribution for a fixed disorder realization by $P_J(q)$ and its disorder average by $P(q)=\mathbb{E}_J[P_J(q)]$. The individual-instance comparisons below concern $P_J(q)$, whereas the small-overlap statistic in Fig.~\ref{fig:4}e is averaged over disorder.

\subsubsection{Sherrington-Kirkpatrick model}
\begin{figure}[h!]
	\begin{center}
		\includegraphics[width=\columnwidth]{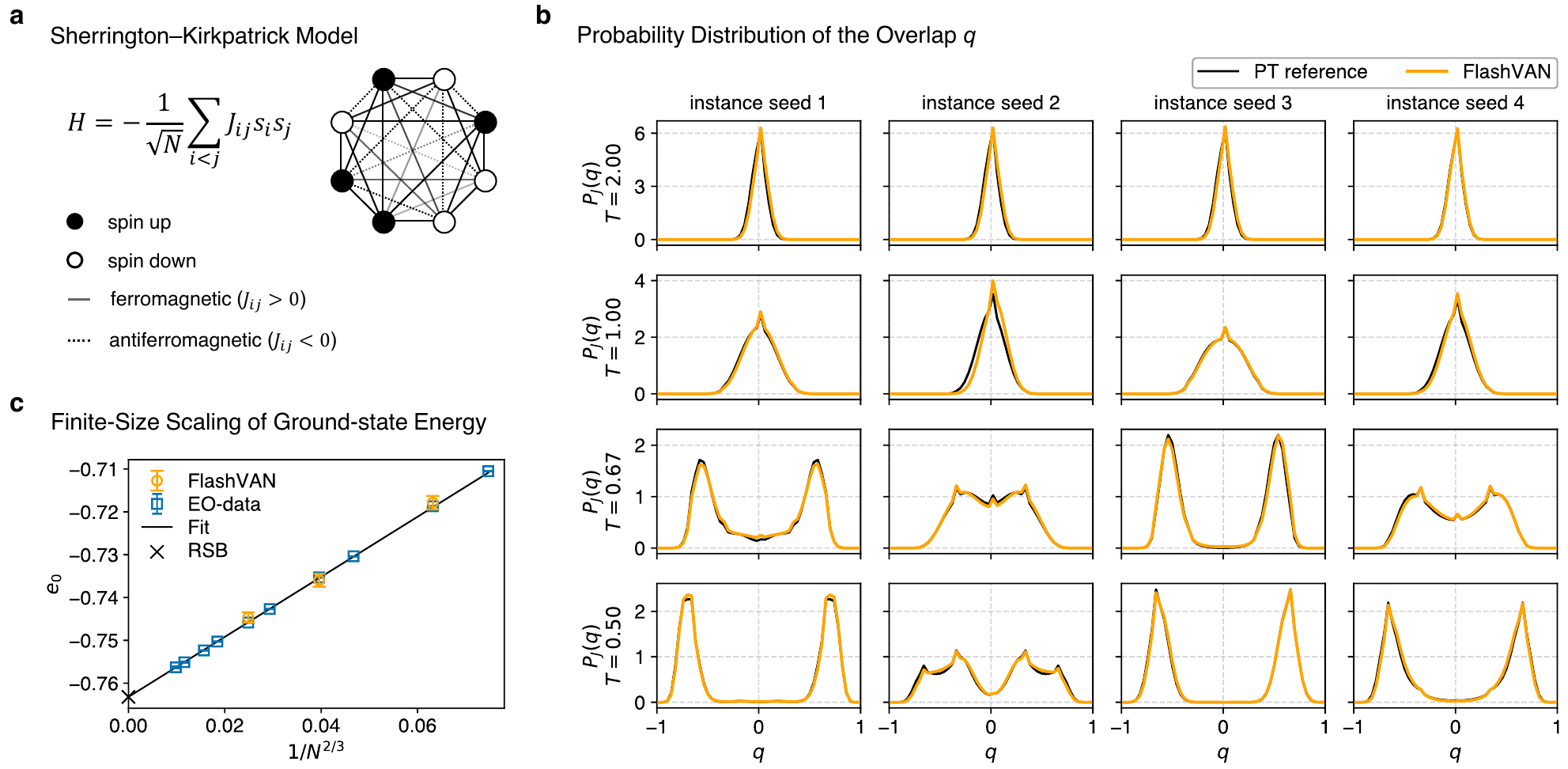}
	\end{center}
	\caption{\textbf{Overlap and ground-state energy benchmarks for the Sherrington-Kirkpatrick (SK) model.} (a) Schematic illustration of the fully connected SK model. (b) Fixed-disorder overlap distributions $P_J(q)$ for Gaussian couplings at $N=256$ and $T=2,1,2/3,1/2$, comparing FlashVAN with parallel tempering (PT) simulations~\cite{responsePT1996} on the same disorder instances. (c) Finite-size scaling of the disorder-averaged ground-state energy estimates for equiprobable couplings $J_{ij}=\pm1$. FlashVAN results are obtained for $N=63,127,255$ over 200, 100, and 50 disorder instances, respectively, and compared with published extremal optimization (EO) data~\cite{responseEO2005}. Error bars represent standard errors across disorder instances. The solid line represents the literature finite-size scaling reference $e_0(N)=-0.76321+0.701N^{-2/3}$~\cite{boettcher2023deep}, while the cross indicates the thermodynamic-limit ground-state energy from Parisi's replica symmetry breaking (RSB) solution~\cite{Mzard1987SpinGT}.}
	\label{fig:2}
\end{figure}

The SK model is a classical benchmark for spin-glass behavior, defined on $N$ binary spins $\boldsymbol{\sigma}=(\sigma_1,\dots,\sigma_N)$ with $\sigma_i\in \left\{ \pm 1 \right\}$. Its Hamiltonian is
\begin{equation}\label{eq:SKHam}
    H\left( \boldsymbol{\sigma} \right) =-\frac{1}{\sqrt{N}}\sum_{i<j}{J_{ij}\sigma_i \sigma_j},
\end{equation}

where $J_{ij}=J_{ji}$ and $J_{ii}=0$. For the overlap experiments, the couplings $J_{ij}$ ($i<j$) are independently drawn from a standard Gaussian distribution. For the ground-state energy comparison, we instead use symmetric bimodal couplings $J_{ij}=\pm1$ with equal probability, matching the EO benchmark. Both use the normalization in Eq.~\ref{eq:SKHam}.

Figure~\ref{fig:2}b compares the fixed-disorder overlap distributions $P_J(q)$ obtained by FlashVAN with parallel tempering (PT)~\cite{responsePT1996} at $N=256$ and $T=2,1,2/3,1/2$, spanning the paramagnetic, critical, and spin-glass regimes. The study covers ten disorder instances, of which four are shown in the main text and six in Supplementary Fig.~8. For each comparison, FlashVAN generates 20,000 configurations directly from the trained model, without additional Monte Carlo refinement or explicitly imposing spin-flip symmetry. The PT reference uses the same couplings, system size, and temperature. FlashVAN reproduces the instance-dependent overlap structures, including the distinct low-temperature peaks, in close agreement with the PT benchmarks. Beyond energy agreement alone, the close correspondence with PT overlap distributions provides strong evidence that FlashVAN captures nontrivial finite-temperature equilibrium statistics across different temperature regimes.

For disorder ensembles such as SK and EA, the disorder-averaged ground-state energy density admits a well-defined thermodynamic limit as $N \rightarrow \infty$. In practice, this limit is approached in a systematic way through finite-size corrections (FSC). Denoting the thermodynamic-limit value by $\langle e_0 \rangle _{N=\infty} $, a commonly used asymptotic form is~\cite{boettcher2023deep}:
\begin{equation}\label{eq:FSC}
    \langle e_0 \rangle _{N} \sim \langle e_0 \rangle _{N=\infty} + \frac{A}{N^\omega} + \dots, N \rightarrow \infty,
\end{equation}
where $A$ is a constant and $\omega(>0)$ is the correction exponent. While additional subleading terms or alternative correction forms may be present, checking whether finite-$N$ data are consistent with this scaling provides a useful baseline for assessing how well a heuristic method extrapolates toward the large-$N$ regime.

Figure~\ref{fig:2}c examines the finite-size scaling of the SK ground-state energy alongside published extremal optimization (EO) results~\cite{responseEO2005}. For the SK model, the asymptotic form in Eq.~\ref{eq:FSC} is characterized by a correction exponent $\omega=2/3$, with the literature values $\langle e_0\rangle_{N=\infty}=-0.76321$ and $A=0.701$~\cite{boettcher2023deep}. The corresponding scaling curve is taken directly from the literature rather than fitted to our data. We evaluate FlashVAN at $N=63,127,255$ over 200, 100, and 50 independent disorder instances, respectively, using symmetric bimodal couplings to match the EO benchmark. For each disorder instance, the energy density is estimated by averaging the energies per spin of 20,000 configurations sampled directly from the trained model, following annealing to $T=0.01$ and fixed-temperature refinement. The disorder-averaged energy densities agree with the EO results within approximately one standard error of the FlashVAN estimates and follow the expected finite-size scaling trend. The cross indicates the thermodynamic-limit ground-state energy predicted by Parisi's replica symmetry breaking (RSB) solution~\cite{Mzard1987SpinGT}, providing a theoretical reference for the large-system limit.

We further benchmark FlashVAN against incremental approximate message passing (IAMP), an algorithm associated with theoretical guarantees in the large-system limit~\cite{responseMontanariSK}, using its publicly available implementation~\cite{responseIAMPImplementation2020}. On 30 matched Gaussian SK instances at $N=256$, FlashVAN achieves lower best-configuration energies in 22 cases, with the remaining eight agreeing within numerical precision. This comparison provides additional evidence of competitive finite-size optimization performance, while the paired results and IAMP settings are presented in Supplementary Table~9. Together with the EO benchmark and finite-size scaling results, this comparison further supports the effectiveness of FlashVAN for ground-state optimization in the SK model.

\subsubsection{Edwards-Anderson model in 2D}

\begin{figure}[!t]
	\begin{center}
		\includegraphics[width=\columnwidth]{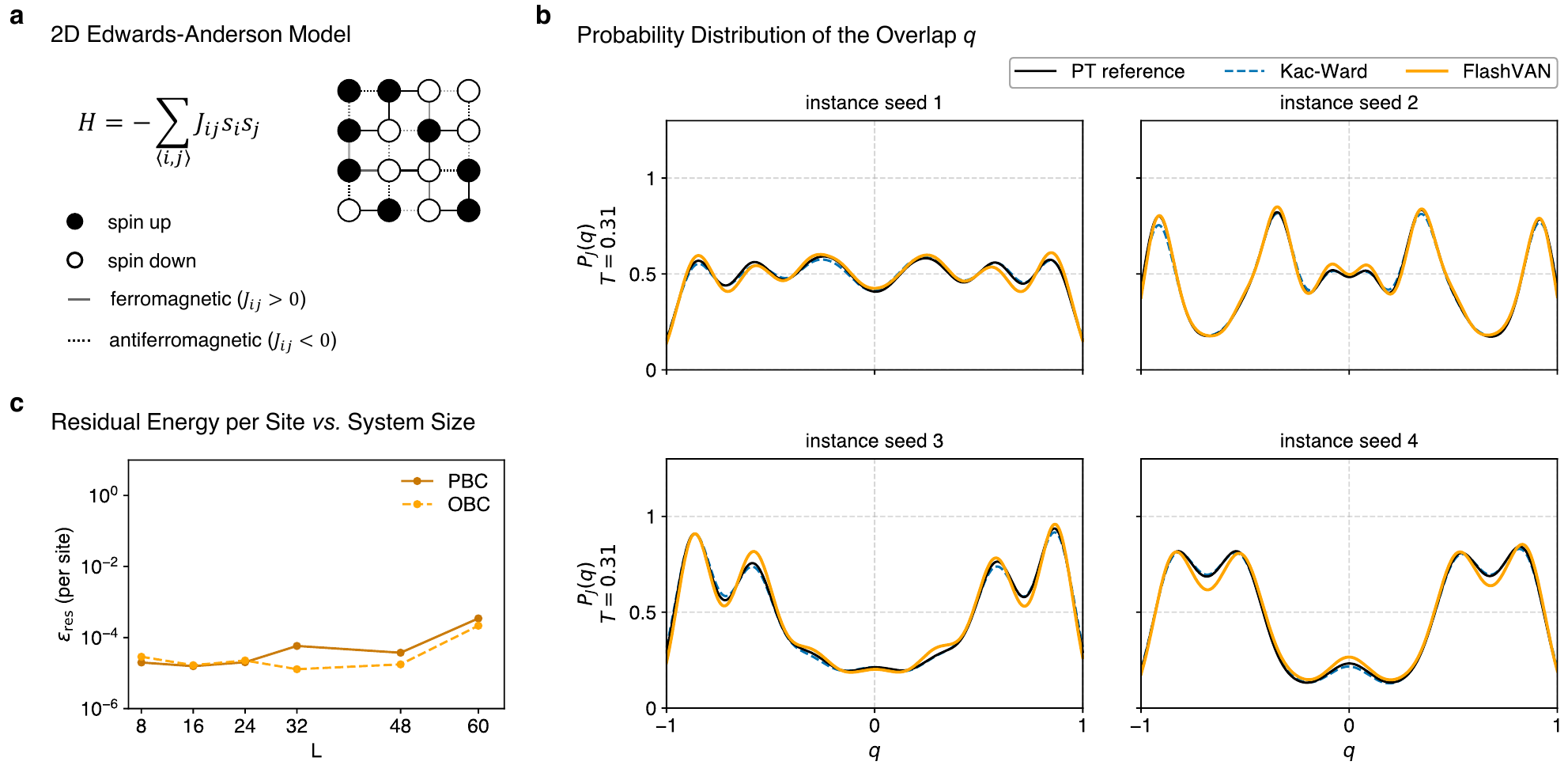}
	\end{center}
	\caption{\textbf{Finite-size overlap and energy benchmarks for the 2D Edwards-Anderson (EA) model.} (a) Schematic illustration of the square-lattice EA model with nearest-neighbor interactions. (b) Fixed-disorder overlap distributions $P_J(q)$ for four $L=16$ instances with open boundary conditions at $T=0.3125$, compared with parallel tempering (PT)~\cite{responsePT1996} and the exact autoregressive Kac-Ward sampler~\cite{responseKacWardSampler2026,PhysRev.88.1332}. (c) Geometric mean of the absolute energy residual per spin relative to McGroundState~\cite{CJMM22} as a function of the linear system size $L$, for open (OBC) and periodic (PBC) boundary conditions.}
	\label{fig:3}
\end{figure}

The EA Ising spin-glass model is defined in general dimension $D$ by
\begin{equation}\label{eq:EAHam}
    H = -\sum_{\langle ij \rangle}J_{ij}\sigma_i\sigma_j,    
\end{equation}
where $\sigma_i\in\{\pm1\}$ are Ising spins, and the summation $\langle ij\rangle$ runs over all nearest-neighbor pairs on a $D$-dimensional lattice. The couplings $J_{ij}$ are independent Gaussian random variables with zero mean and unit variance. For Gaussian disorder, the ground state is almost surely unique up to a global spin flip for any finite system. Each specific realization of $\{J_{ij}\}$ is referred to as a disorder instance. For the 2D overlap comparison, we use the fixed disorder instances provided by Del Bono et al.~\cite{del2025nearest}, while the energy benchmarks use independently generated Gaussian couplings. In the following, we first present results for the 2D case, where the system size is $N=L^2$; results for the 3D case will be discussed in the next section.

Figure~\ref{fig:3}b compares the fixed-disorder overlap distributions $P_J(q)$ for four $L=16$ instances with open boundary conditions at $T=0.3125$. FlashVAN is benchmarked against parallel tempering (PT)~\cite{responsePT1996} and the exact autoregressive sampler of Liu et al.~\cite{responseKacWardSampler2026}, based on the Kac-Ward construction~\cite{PhysRev.88.1332}. FlashVAN generates 200,000 configurations directly from the trained model for each instance. The close agreement across the distinct multipeaked distributions demonstrates FlashVAN's ability to reproduce nontrivial finite-size equilibrium overlap statistics for individual disorder realizations. Importantly, these features do not imply a finite-temperature spin-glass transition, consistent with the established $T_c=0$ behavior of the 2D Gaussian EA model~\cite{responseHoudayerHartmann2004}.

We next evaluate the ground-state energy accuracy of FlashVAN across system sizes $L=8,16,24,32,48,60$ under both open and periodic boundary conditions. For each disorder instance, the predicted energy density $e^{\mathrm{pred}}$ is obtained from the final mean energy per spin and compared with the exact ground-state energy density $e_0^{\mathrm{exact}}$ computed using McGroundState~\cite{CJMM22}. Figure~\ref{fig:3}c presents the geometric mean of the absolute residual $\epsilon_{\mathrm{res}}=|e^{\mathrm{pred}}-e_0^{\mathrm{exact}}|$ across disorder instances. The residuals remain approximately within $10^{-5}$-$10^{-4}$ for most system sizes, increasing to a few $10^{-4}$ at $L=60$. These results demonstrate that FlashVAN maintains high energy accuracy across a broad range of lattice sizes and boundary conditions, with its predictions closely approaching the exact ground-state energies.

Recent Transformer-based neural samplers have also been applied to 2D EA systems of comparable size, reaching $64\times64$ spins in finite-temperature Boltzmann-distribution sampling~\cite{bialas2026sampling}. FlashVAN complements these developments by addressing both finite-temperature overlap statistics and low-temperature ground-state energy estimation, while its computational efficiency is evaluated separately in Fig.~\ref{fig:1}e and Supplementary Table~5.

\subsubsection{Edwards-Anderson model in 3D}

\begin{figure}[!t]
	\begin{center}
		\includegraphics[width=\columnwidth]{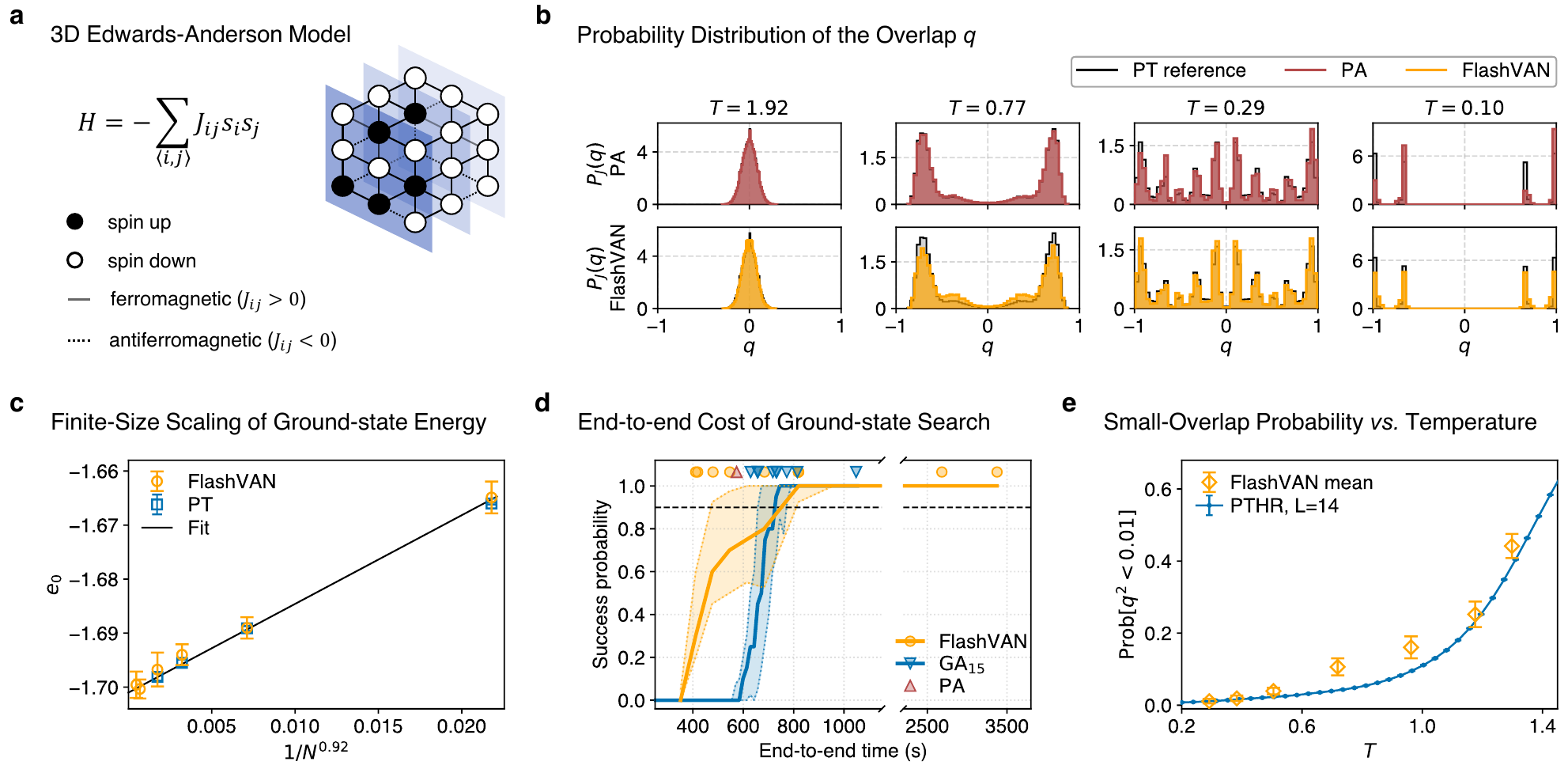}
	\end{center}
	\caption{\textbf{Overlap, energy, and computational-cost benchmarks for 3D EA.} (a) Cubic-lattice EA interactions. (b) $P_J(q)$ for one $L=10$ instance at four temperatures, comparing FlashVAN and PA with PT data from Ref.~\cite{del2026demonstrating}. (c) Disorder-averaged energy versus size, with PT references~\cite{ROMA20092821,responseWangGroundStates2015} and the literature finite-size scaling reference $-1.701+1.641N^{-0.92}$~\cite{boettcher2023deep}. Error bars are disorder standard errors. (d) Ground-state-search success probability versus time for ten matched $L=14$ instances: median curves and interquartile bands, with individual attainment-budget markers. The GA curve uses hardware-calibrated published times~\cite{del2026demonstrating}; GA/PA markers use RTX~5090 reruns. (e) Disorder-averaged $\Pr_J(q^2<0.01)$ at $L=14$ compared with parallel tempering enhanced with Houdayer moves and an entropic reservoir (PTHR)~\cite{chilin2026true,responseChilinPTHR2026}: 60 instances at each of the three lower temperatures and ten at each of the four higher temperatures. FlashVAN errors are disorder-bootstrap standard errors.}
	\label{fig:4}
\end{figure}

We next consider the more challenging 3D EA model. Figure~\ref{fig:4}b compares the fixed-disorder overlap distributions $P_J(q)$ obtained by FlashVAN with the parallel tempering (PT) reference for an $L=10$ instance at four temperatures, $T=1.92,0.77,0.29,0.10$. Population annealing (PA) results from the same benchmark are also included for comparison~\cite{del2026demonstrating}. FlashVAN closely reproduces the temperature-dependent evolution of the overlap distributions, from the central peak at high temperature to the increasingly complex, multipeaked structures at lower temperatures. The agreement with PT across these distinct regimes provides further evidence that FlashVAN can capture nontrivial finite-temperature equilibrium statistics beyond energy-related observables. For consistency with the reference data, the plotted FlashVAN and PT overlap histograms are symmetrized under global spin reversal.

 To further assess the scalability of FlashVAN, we perform a finite-size scaling analysis of the ensemble-averaged ground-state energy density, using the finite-size scaling form in Eq.~\ref{eq:FSC}. Figure~\ref{fig:4}c presents the results for $L=4,6,8,10,14,16$, averaged over 1110, 651, 300, 50, 40, and 20 disorder instances, respectively. The solid line represents the numerical finite-size scaling reference $-1.701+1.641N^{-0.92}$ reported by Boettcher~\cite{boettcher2023deep}, corresponding to Eq.~\ref{eq:FSC} with $\langle e_0\rangle_{N=\infty}=-1.701$, $A=1.641$, and $\omega=0.92$. For comparison, we also include the available parallel tempering (PT) benchmarks of Rom{\'a} et al.~\cite{ROMA20092821}, also tabulated by Wang et al.~\cite{responseWangGroundStates2015}. The ground-state energy estimates obtained by FlashVAN agree well with both the numerical finite-size scaling reference and the independent PT benchmarks, supporting its accuracy across the investigated system sizes.

Figure~\ref{fig:4}d presents an end-to-end computational-cost comparison between FlashVAN, global annealing (GA), and population annealing (PA) on ten matched $L=14$ disorder instances from Del Bono et al.~\cite{del2026demonstrating}. For FlashVAN, the measured time includes pretraining, annealing, and refinement, while initial high-temperature population preparation is excluded following the GA benchmark convention. A run is considered successful when its best energy agrees with the reported reference to five decimal places. We characterize the search cost using the attainment time $t_{0.9}$, defined as the mean runtime at the first tested budget achieving a success rate of at least 90\%. The median $t_{0.9}$ is $752$~s for FlashVAN and $724$~s for GA, with each method achieving shorter attainment times on five of the ten instances. These results demonstrate comparable median search costs and complementary instance-dependent advantages with training included. For example, on instance 502033, FlashVAN reaches the success threshold in $419$~s compared with $717$~s for GA. The contrasting performance across matched disorder instances may reflect different optimization bottlenecks. This instance dependence motivates further investigation of whether their complementary strengths can be exploited in hybrid optimization strategies. The same framework also supports finite-temperature distribution learning. The GA curves are based on published V100 timings calibrated to RTX~5090 performance, whereas the GA and PA markers correspond to directly timed RTX~5090 reruns. The timing convention is described in Methods, and per-instance results are listed in Supplementary Table~11.

To examine the contribution of the training strategy beyond the Transformer backbone, we replace the autoregressive model with the simpler MADE architecture used in GA~\cite{germain2015made,del2026demonstrating}, using the same population-assisted training strategy and evaluation protocol. On instance 11086240, which has the longest GA attainment time in the ten-instance benchmark, MADE with our training strategy achieves ten successes out of ten runs at the first tested budget reaching the 90\% threshold. Its attainment time is approximately $343$~s, compared with $1048$~s for GA and $3375$~s for Transformer-based FlashVAN, giving a $3\times$ speedup relative to GA under the same benchmark convention (Supplementary Fig.~11 and Table~12). This result highlights the potential of population-assisted training beyond a specific neural architecture. For the broader objectives of this work, we retain the Transformer backbone to support finite-temperature distribution learning and larger-system modeling. Its flexible attention-based architecture also provides a natural foundation for future extensions, such as disorder-conditioned generation, while the MADE variant merits further investigation for ground-state optimization.

Agreement in energy alone does not establish whether a sampling method correctly reproduces the relative statistical weights of weakly overlapping configurations. We therefore examine the disorder-averaged small-overlap probability
\begin{equation}\label{eq:small_overlap_weight}
I_L(T)=\mathbb{E}_J\!\left[\Pr_J(q^2<0.01)\right]
      =\int_{-0.1}^{0.1}P(q;L,T)\,dq,
\end{equation}
which probes the probability of sampling configurations with small mutual overlap and provides a complementary measure of low-temperature spin-glass statistics. Figure~\ref{fig:4}e compares FlashVAN with the independent PTHR results of Chilin et al.~\cite{chilin2026true,responseChilinPTHR2026}. At $L=14$, we consider 60 disorder instances at each of the three lower temperatures and ten at each of the four higher temperatures, totaling 220 instance-temperature settings. For each setting, FlashVAN generates 20,000 configurations directly from the trained model, yielding 10,000 replica-pair overlaps without additional Monte Carlo refinement. At the two lowest temperatures, FlashVAN obtains $I_{14}(T)=0.0110\pm0.0063$ and $0.0189\pm0.0077$, compared with the PTHR reference values of $0.0103$ and $0.0162$, respectively. The uncertainties are disorder-bootstrap standard errors. At these two lowest temperatures, the nonzero small-overlap probabilities agree with the independent reference within the reported uncertainties, providing a more sensitive test of low-temperature sampling than energy agreement alone. The two methods use independent disorder ensembles, and the PTHR comparisons employ the nearest published temperatures. Additional results for $L=16$, covering ten disorder instances at each of seven temperatures and showing individual-instance small-overlap probabilities, are provided in Supplementary Fig.~9.

\section{Discussion}

We have developed FlashVAN, an autoregressive framework that integrates physics-inspired attention, concatenated positional embeddings, population-assisted training, and hardware-efficient acceleration for spin-glass systems. Across the Sherrington-Kirkpatrick and Edwards-Anderson models, FlashVAN demonstrates competitive ground-state optimization and the ability to reproduce nontrivial finite-temperature overlap statistics. A key aspect of the framework is the combination of Monte Carlo exploration with autoregressive learning, allowing the model to incorporate information from a refined sample population while retaining a directly sampleable probability representation. This integration extends the practical capabilities of neural autoregressive approaches to disordered systems.

The practical value of FlashVAN lies in its computational efficiency and the capabilities of its learned generative representation. Compared with GA, a highly competitive machine-learning-assisted Monte Carlo approach, FlashVAN achieves comparable median ground-state-search costs across the matched benchmarks, with complementary advantages on different disorder instances. This instance dependence suggests opportunities to combine the strengths of both approaches. Applying our population-assisted training strategy to the simpler MADE architecture substantially accelerates the instance with the longest GA attainment time, demonstrating that the training approach can also be effective beyond the Transformer backbone. While simpler architectures may be advantageous for particular optimization tasks, we retain the Transformer-based framework for its broader applicability to finite-temperature distribution learning and its flexibility for future extensions. In particular, recent progress in disorder-conditioned generation through cross-attention~\cite{responseUniversalSampling2026} suggests a route toward reusable generative models across disorder realizations. The ability to produce independent configurations from the learned distribution with evaluable probabilities, alongside single-GPU simulations of systems with up to $4096$ spins, further illustrates the potential of FlashVAN as a versatile generative framework for spin-glass investigations.

The ability to reproduce overlap statistics also connects FlashVAN to fundamental questions about the low-temperature organization of spin glasses. In particular, the overlap distribution $P_J(q)$ and the disorder-averaged low-overlap weight $I_L(T)$ characterize aspects of the statistical structure of equilibrium configurations that cannot be inferred from energy alone. For the three-dimensional Edwards-Anderson model, the size and temperature dependence of low-overlap weights is relevant to the longstanding debate between replica-symmetry-breaking and droplet descriptions of the spin-glass phase. For 3D EA at $L=14$, our results reproduce small but nonzero low-overlap weights at the two lowest investigated temperatures, consistent with independent Monte Carlo references within the reported uncertainties. Recent large-scale simulations~\cite{chilin2026true} provide valuable benchmarks for further investigating how $I_L(T)$ evolves with system size, temperature, and disorder realization. Such systematic studies may provide additional numerical evidence relevant to competing theoretical descriptions.

Several challenges remain for applying FlashVAN to broader spin-glass investigations. The accuracy of the learned distribution depends on system size, temperature, and disorder realization, particularly in the low-temperature regime where sampling across competing configurations becomes increasingly difficult. Although autoregressive generation produces independent configurations, these samples may retain systematic biases from imperfect population exploration and generative learning. Moreover, the present framework requires separate training for each disorder realization and target temperature, making computational cost an important consideration for extensive disorder-averaged studies. The reported numerical agreement does not constitute a rigorous guarantee of equilibration or asymptotic correctness. In particular, the low-temperature SK sampling obstruction established by El Alaoui, Montanari, and Sellke~\cite{responseSKSampling2022} applies to algorithms satisfying a specific stability condition under disorder perturbations; our finite-size observable tests neither establish that FlashVAN satisfies this condition nor demonstrate that it evades the obstruction. Improving population exploration, generative accuracy, and transferability could further strengthen neural generative models as complementary tools for investigating unresolved problems in spin-glass physics.

\section{Methods}

\subsection{\texorpdfstring{Model architecture}{Model architecture}}
\label{sec:model_architecture}
In this section, we present the pseudocode (Algorithm~\ref{alg:flashvan}) outlining the implementation of FlashVAN, as described in the Results section. Compared with conventional decoder-only transformer architectures, our design differs in two key aspects: the physics-inspired attention mechanism and the concatenated positional embedding. The former exploits the interaction structure of spin-glass systems to improve computational efficiency, while the latter  enhances the expressive capacity of FlashVAN.

\begin{algorithm}[H]
  \begin{algorithmic}[1]
       \State Input configuration $\boldsymbol{\sigma} \in \{-1,+1\}^{N}$
      \State Concatenate Positional Embed: $\mathcal{X} \leftarrow (\mathbf{x}_1, \dots \mathbf{x}_{N}) \in \mathbb{R}^{d}$
      \For{$i = 1, n_l$} \label{forloop}
        \State $\mathcal{X} \leftarrow \mathcal{X} + \text{MHA}(\text{LayerNorm($\mathcal{X}$)})$ \Comment{physics-inspired attention}
        \State $\mathcal{X} \leftarrow \text{LayerNorm($\mathcal{X}$)}$
        \State $\mathcal{X} \leftarrow \mathcal{X} + \text{FFN}(\mathcal{X})$
      \EndFor
      \State $(\mathbf{z}_1, \dots \mathbf{z}_{N}) \leftarrow \text{OutputHead}(\mathcal{X})$
      \State $\log[q(\boldsymbol{\sigma})] \leftarrow \sum_{i=1}^N \log(\text{softmax}(\mathbf{z}_i)_{\sigma_i})$
  \end{algorithmic}
  \caption{FlashVAN} \label{alg:flashvan}
\end{algorithm}

\subsubsection{\texorpdfstring{Interaction structure and attention patterns}{Interaction structure and attention patterns}}
\label{sec:attention_patterns}
Consider an Ising model defined on a graph $G=(V,E)$ with $|V| = N$ spins, where $E \subseteq \{(i, j) : i, j \in V, i \neq j\}$ denotes the set of edges encoding pairwise interactions. The Ising model naturally defines a Markov random field (MRF) whose joint distribution $P(\boldsymbol{\sigma})$ follows the Boltzmann distribution. The \emph{local Markov property} of the MRF states that each spin $\sigma_i$ is conditionally independent of all other spins given its immediate neighbors (its \emph{Markov blanket}).

This interaction-based conditional structure has also been exploited in TwoBo~\cite{Biazzo_Wu_Carleo_2024} and tensor-network Monte Carlo methods~\cite{frias-perez2023collective,chen2025tnmc,chen2025batchtnmc}. While TwoBo uses coupling-weighted effective fields and a feed-forward network, FlashVAN incorporates the interaction structure into its attention mechanism.

To enable autoregressive sampling we impose a fixed total ordering $\prec$ on the vertex set $V$ (for example, a raster scan). This ordering gives an exact factorization of the joint distribution via the chain rule:
\begin{equation}
    P(\boldsymbol{\sigma}) = \prod_{i=1}^N P(\sigma_i \mid \boldsymbol{\sigma}_{\prec i}),
\end{equation}
where $\boldsymbol{\sigma}_{\prec i}$ denotes the configuration of all spins preceding $i$ under the ordering. Let $P_i = \{j \in V : j \prec i\}$ be the set of ``past'' (already sampled) spins and $F_i = \{j \in V : j \succ i\}$ the set of ``future'' (not yet sampled) spins. Because of the Markov properties of the underlying graph, the autoregressive conditional $P(\sigma_i \mid \boldsymbol{\sigma}_{P_i})$ need not depend on the entire past $P_i$. One can therefore identify a minimal conditioning set, which we call the \emph{ordered Markov boundary}.

Given the ordering $\prec$, the ordered Markov boundary of spin $i$, denoted $\partial_{\prec}(i)$, is the minimal subset of $P_i$ such that $\sigma_i$ is conditionally independent of the remaining past given $\boldsymbol{\sigma}_{\partial_{\prec}(i)}$:
\begin{equation}\label{eq:boundary_conditional}
\sigma_i \perp\!\!\!\perp \boldsymbol{\sigma}_{P_i \setminus \partial_{\prec}(i)} \mid \boldsymbol{\sigma}_{\partial_{\prec}(i)} .
\end{equation}
Graphically, $\partial_{\prec}(i)$ consists of those past nodes $j\in P_i$ that are either directly adjacent to $i$ in $G$, or connected to $i$ by a path $j \rightarrow v_1 \rightarrow \cdots \rightarrow v_k \rightarrow i$, whose intermediate nodes $v_1,\dots,v_k$ all lie in the unsampled future $F_i$. Intuitively, only past nodes that can influence $i$ through chains passing entirely via future nodes remain relevant after marginalizing out $F_i$.

Consequently, the autoregressive conditional depends only on the ordered Markov boundary: $P(\sigma_i \mid \boldsymbol{\sigma}_{P_i}) = P(\sigma_i \mid \boldsymbol{\sigma}_{\partial_{\prec}(i)})$. An explicit expression for the conditional is obtained by marginalizing over the future variables $F_i$:
\begin{equation}
P(\sigma_i \mid \boldsymbol{\sigma}_{P_i}) = \frac{\sum_{\boldsymbol{\sigma}_{F_i}} P(\boldsymbol{\sigma}_{P_i}, \sigma_i, \boldsymbol{\sigma}_{F_i})}{\sum_{\sigma_i, \boldsymbol{\sigma}_{F_i}} P(\boldsymbol{\sigma}_{P_i}, \sigma_i, \boldsymbol{\sigma}_{F_i})}.
\end{equation}

In graphical-model language, marginalizing out a set of nodes $F_i$ induces a new graph structure over the remaining nodes $P_i \cup \{i\}$. Specifically, marginalization fully connects the neighbors of any marginalized node, effectively creating ``fill-edges'' across the boundary separating $P_i$ and $F_i$. Therefore, the only variables in $P_i$ that remain directly dependent on $\sigma_i$ after marginalizing $F_i$ are those connected to $i$ via paths through $F_i$, or those directly adjacent to $i$. This set is exactly $\partial_{\prec}(i)$.

This observation guides the construction of sparse causal attention: the conditioning inputs must retain the ordered Markov boundary. The dependence structure is exact, whereas the neural approximation to each conditional probability is assessed numerically. For the fully connected SK model, the ordered Markov boundary contains all preceding spins, and full causal attention is therefore used.

\subsubsection{\texorpdfstring{Positional representations}{Positional representations}}
\label{sec:concatePE}
Unlike natural language sequences, spin configurations have a limited local state space, while the spatial arrangement and interactions between spins carry essential physical information. This motivates our concatenated positional embedding, which preserves spin-state and positional features in separate dimensions rather than combining them through addition.

Specifically, let $d_{\text{token}}$ and $d_{\text{pos}}$ denote the token and positional embedding dimensions, respectively, with the overall embedding dimension $d = d_{\text{token}}+d_{\text{pos}}$. For the token at position $i$, the input representation is constructed as follows:
\begin{equation}\label{eq:concatePE}
    \begin{aligned}
        \mathbf{s}_i &= \mathrm{Embed}(\sigma_i) \in \mathbb{R}^{d_{\text{token}}}, \\[4pt]
        \mathbf{p}_i &= \mathrm{Embed}(p_i) \in \mathbb{R}^{d_{\text{pos}}}, \\[4pt]
        \mathbf{x}_i &= [\,\mathbf{s}_i \,;\, \mathbf{p}_i\,] \in \mathbb{R}^{d}.
    \end{aligned}
\end{equation}
The operator $[\,\cdot\,;\,\cdot\,]$ denotes vector concatenation. We allocate more embedding dimensions to positional features than to spin-state features, emphasizing positional information in the input representation.

For patch-based tokenization, the state embedding represents the joint spin configuration within each patch rather than an individual spin. A $2\times2$ patch has 16 possible states, while a $2\times2\times2$ patch has 256, reducing the sequence lengths to $(L/2)^2$ and $(L/2)^3$, respectively.

For the EA models, we additionally employ spatial rotary positional embeddings (RoPE)~\cite{RoPE}, which encode relative positional information by rotating the attention queries and keys. The rotations are based on the integer coordinates of patches on the coarse lattice, with separate feature groups assigned to each spatial axis. RoPE thus complements the concatenated input embedding, while patching reduces the sequence length. The SK model uses single-spin tokens with concatenated positional embeddings, without RoPE or patching. Axis allocations and rotation frequencies are detailed in Supplementary Section~I.C; positional-encoding comparisons are shown in Supplementary Figs.~3-5. Embedding dimensions for the energy benchmarks and model-specific components are listed in Supplementary Tables~2-8.

\subsection{\texorpdfstring{Training procedure}{Training procedure}}
\label{sec:TrainingStrategy}

\subsubsection{\texorpdfstring{Training workflow}{Training workflow}}
\label{sec:training_workflow}
The population-assisted training strategy introduced in Results uses a persistent sample pool to train FlashVAN (Fig.~\ref{fig:1}b). The pool is initialized from randomized configurations and subsequently replenished with autoregressively generated samples. Pool configurations are refined through standard local Metropolis Monte Carlo (LMC) updates~\cite{Metropolis1953}: single-spin-flip moves accepted with probability $\min\{1,\exp(-\Delta E/T)\}$, where $\Delta E$ is the energy change of the proposed flip. One Monte Carlo sweep (MCS) consists of one attempted update per spin per configuration. After initialization, training comprises three stages: \textbf{annealing, refinement, and distillation}.

\textbf{Initialization.} We first relax a large randomly initialized population using LMC at a high temperature $T_0$. FlashVAN is then pretrained on this population through maximum-likelihood learning, and a uniformly selected random subset initializes the smaller persistent sample pool. Detailed Stage~0 operations and the subsequent three stages are described in Supplementary Section~II.B.

\textbf{Annealing.} The model and pool co-evolve as the temperature decreases from $T_0$ to the target temperature. When the temperature changes from $T_{k-1}$ to $T_k$, pool configurations are reweighted by $\exp[-(\beta_k-\beta_{k-1})E]$, where $\beta_k=1/T_k$, followed by resampling at fixed pool size. Fresh model-generated configurations replace a subset of the pool, followed by LMC updates at $T_k$. FlashVAN is then updated using batches drawn from the refined pool, according to the training objective defined in the next subsection. 

\textbf{Refinement.} At the target temperature, model-sample injection and population resampling are suspended. The pool continues to relax through LMC, while FlashVAN continues to learn from pool batches using the same training objective. In the finite-temperature experiments, the duration is determined by stagnation of a pool-based self-consistency diagnostic. For centered unit-variance Gaussian EA couplings, Gaussian integration by parts gives the disorder-averaged equilibrium relation
\begin{equation}\label{eq:gaussian_pool_identity}
\mathbb E_J\!\left[\frac{\langle E\rangle_J}{B}+\frac{1-\langle q_\ell\rangle_J}{T}\right]=0,\qquad
q_\ell=\frac{1}{B}\sum_{\langle ij\rangle}\sigma_i^1\sigma_j^1\sigma_i^2\sigma_j^2,
\end{equation}
where $B$ is the number of interacting bonds and $q_\ell$ is the link overlap between two replicas with the same disorder~\cite{chilin2026true}. For a periodic cubic lattice, $B=3N$, and the pool residual is $S(t)=e_{\mathrm{pool}}(t)/3+[1-q_{\ell,\mathrm{pool}}(t)]/T$. The stopping rule monitors $|S(t)-S(t/4)|$ against a tolerance at consecutive probes, subject to minimum and maximum durations. The SK Gaussian counterpart uses $S=e_{\mathrm{pool}}+(1-\langle q^2\rangle_{\mathrm{pool}})/(2T)$; the fixed-instance 2D protocol uses a generalized bond-moment diagnostic. 

\textbf{Distillation.} The refined pool is frozen, and the same FlashVAN model continues training through pure maximum-likelihood learning using configurations drawn from this pool. No further LMC updates or model-sample injection are performed, and the temperature remains fixed. Distillation transfers the refined population statistics into the autoregressive model. Subsequent measurements use independent configurations generated directly from the trained model, as described in Methods Sec.~\ref{sec:direct_sampling}.

We apply population-assisted training across the SK, 2D EA, and 3D EA models. Population-assisted task protocols and LMC parameters are summarized in Supplementary Tables~2, 4, and~7. The energy-scaling datasets in Figs.~\ref{fig:3}c and~\ref{fig:4}c retain the separate annealing configurations in Supplementary Tables~6 and~8. For ground-state search, annealing approaches a low terminal temperature, and refinement, when used, monitors stagnation of the best pool energy. The choice of training stages depends on the task. Finite-temperature overlap measurements use direct model samples.

Muon~\cite{jordan2024muon,liu2025muon} is used for matrix-valued parameter updates. Optimizer settings and convergence comparisons are reported in Supplementary Tables~2, 4, and~7 and Supplementary Fig.~6.

\subsubsection{\texorpdfstring{Training objective}{Training objective}}
\label{sec:training_objective}
FlashVAN learns a normalized autoregressive distribution $q_\theta$ to approximate the target Boltzmann distribution. At temperature $T$, the underlying variational free-energy objective satisfies~\cite{PhysRevLett.122.080602}
\begin{equation}\label{eq:KL_div}
D_{\mathrm{KL}}(q_\theta\|p)=\beta(F_q-F)\geq0,
\end{equation}
\begin{equation}\label{eq:free_energy}
F_q=\mathbb E_{q_\theta}\!\left[E(\boldsymbol\sigma)+T\ln q_\theta(\boldsymbol\sigma)\right],\qquad T=\beta^{-1},
\end{equation}
where $F=-T\ln Z$ is the equilibrium free energy. Defining $L_\theta(\boldsymbol\sigma)=E(\boldsymbol\sigma)+T\ln q_\theta(\boldsymbol\sigma)$, with $E$ the total configuration energy, gives the score-function identity~\cite{williams1992simple}
\begin{equation*}
\nabla_\theta F_q=\mathbb E_{q_\theta}\!\left[\left(L_\theta-\mathbb E_{q_\theta}L_\theta\right)\nabla_\theta\ln q_\theta\right].
\end{equation*}
Conventional VAN training estimates this gradient using samples generated directly from $q_\theta$. In contrast, during annealing and refinement, FlashVAN is trained on configurations drawn from the persistent pool, which incorporates information acquired through local Monte Carlo refinement. The model therefore learns from configurations explored and relaxed by the population, beyond its own current sampling distribution.

For a pool-drawn batch $\{\boldsymbol\sigma^{(m)}\}_{m=1}^{M}$, we introduce a population-assisted training loss that combines a centered free-energy-based learning signal with an additional likelihood term:
\begin{equation}\label{eq:Loss}
\mathcal L_{\mathrm{train}}=\frac{1}{M}\sum_{m=1}^{M}\operatorname{sg}\!\left[L_\theta(\boldsymbol\sigma^{(m)})-\overline L\right]\ln q_\theta(\boldsymbol\sigma^{(m)})-\frac{w}{M}\sum_{m=1}^{M}\ln q_\theta(\boldsymbol\sigma^{(m)}).
\end{equation}
The first term uses the batch-centered value of $L_\theta$ to weight the log-probability update. The operator $\operatorname{sg}$ denotes stop-gradient: the sampled configurations and the centered coefficient are held fixed during differentiation, and gradients flow through the multiplying log probability. The baseline $\overline L=M^{-1}\sum_{m=1}^{M}L_\theta(\boldsymbol\sigma^{(m)})$ is recomputed from each batch and is not independently tuned. The second term encourages the model to fit the refined pool configurations through maximum-likelihood learning, with its contribution controlled by the weight $w$. Unlike the batch-derived baseline, $w$ is specified by the model- and task-dependent training protocol, with values listed in Supplementary Tables~2, 4, and~7. Although $\overline L$ and $w$ can be combined algebraically in the coefficient of $\ln q_\theta$, the former centers the batch signal while the latter sets the additional likelihood contribution.

Because the training samples are drawn from the evolving pool rather than directly from $q_\theta$, the resulting gradient is not an unbiased estimator of the standard variational free-energy gradient.

During initialization and distillation, FlashVAN is trained using the pure maximum-likelihood objective, $\mathcal L_{\mathrm{MLE}}=-M^{-1}\sum_m\ln q_\theta(\boldsymbol\sigma^{(m)})$ without the centered score-function term. Initialization uses the high-temperature population, whereas distillation uses the frozen refined pool at the target temperature. Both stages update the same autoregressive model. This final fit transfers the population statistics into the directly sampleable autoregressive representation.

\subsection{\texorpdfstring{Sampling and computational implementation}{Sampling and computational implementation}}
\label{sec:hardware_acceleration}

\subsubsection{\texorpdfstring{Direct autoregressive sampling}{Direct autoregressive sampling}}
\label{sec:direct_sampling}
FlashVAN generates spin configurations sequentially from normalized conditional probabilities. Starting from a fixed initial token, the model predicts a categorical distribution for the next spin or patch token, samples from this distribution, and conditions subsequent predictions on the generated sequence. For patch-based models, the generated tokens are decoded into their constituent spins to recover the full configuration. During training, this direct-sampling procedure provides fresh model-generated configurations for the persistent sample pool, as described in Methods Sec.~\ref{sec:training_workflow}.

Unlike conventional Markov chain Monte Carlo (MCMC) methods, which generate configurations through successive stochastic transitions, FlashVAN produces independent configurations from its learned distribution without requiring a Markov chain between generated samples. After training, this capability enables the direct evaluation of physical observables, including overlap distributions, using independently generated replicas. Local Metropolis Monte Carlo remains part of the training procedure.

\subsubsection{\texorpdfstring{Computational acceleration}{Computational acceleration}}
\label{sec:computational_acceleration}
To improve computational efficiency, FlashVAN integrates FlashAttention-2~\cite{Dao_2024} and key-value (KV) caching. FlashAttention-2 avoids explicitly materializing the full attention matrix, reducing memory overhead and data movement during training. For the 2D EA model (Fig.~\ref{fig:1}e), FlashVAN reduces the per-step runtime from $3015$ to $172$~ms at $N=576$ ($L=24$) and from $13819$ to $380$~ms at $N=1024$ ($L=32$), corresponding to speedups of $17.5\times$ and $36.4\times$ over the traditional VAN implementation, respectively.

During autoregressive generation, conventional transformer implementations repeatedly compute attention over previously generated tokens, resulting in redundant calculations. To avoid this, FlashVAN adopts a KV cache strategy~\cite{pope2023efficiently,kwon2023efficient}, which stores the keys and values of preceding tokens and reuses them at subsequent generation steps. Only the new token is processed at each step, reducing the cumulative attention complexity from $\mathcal O(n^3)$ to $\mathcal O(n^2)$ for full causal attention over $n$ tokens. The sampling procedure is summarized in Algorithm~\ref{alg:flashvan_kvcache}.

\begin{algorithm}[H]
\begin{algorithmic}[1]
    \State \textbf{Input:} Batch size $B$, sequence length $n$, number of layers $n_\ell$, fixed start token $s_{\mathrm{start}}$
    \State Initialize samples $\mathbf{s}$ and KV caches $\mathcal{C}$
    \State Set initial input token $\mathbf{x}_1 \leftarrow s_{\mathrm{start}}$
    \For{$t=1,\ldots,n$}
        \State $\mathbf{h}_t \leftarrow \mathrm{ConcatEmbed}(\mathbf{x}_t,t)$
        \For{$\ell=1,\ldots,n_\ell$}
            \State $(\mathbf{h}_t,\mathcal{C}^{(\ell)}) \leftarrow \mathrm{TransformerBlock}^{(\ell)}(\mathbf{h}_t,\mathcal{C}^{(\ell)})$ \Comment{KV-cached attention}
        \EndFor
        \State $\mathbf{p}_t \leftarrow \mathrm{softmax}(\mathrm{OutputHead}(\mathbf{h}_t))$
        \State $\mathbf{s}_{:,t}\sim\mathrm{Categorical}(\mathbf{p}_t)$
        \State $\mathbf{x}_{t+1}\leftarrow\mathbf{s}_{:,t}$
    \EndFor
    \State $\boldsymbol{\sigma}\leftarrow\mathrm{Decode}(\mathbf{s})$
    \State \textbf{Output:} Spin configurations $\boldsymbol{\sigma}$
\end{algorithmic}
\caption{Autoregressive sampling with KV caching in FlashVAN.}
\label{alg:flashvan_kvcache}
\end{algorithm}

The population-assisted FlashVAN runs use a single NVIDIA RTX~5090 GPU. In our implementation, attention inputs use half precision, while embeddings, feed-forward operations, and energy evaluations retain single precision. The hardware and precision settings for the corresponding experiments are described in Supplementary Section~IV.A; additional implementation benchmarks are shown in Supplementary Fig.~10.

\subsection{\texorpdfstring{Physical observables and evaluation protocols}{Physical observables and evaluation protocols}}
\label{sec:benchmark_protocols}

To systematically evaluate FlashVAN, we consider multiple physical observables and numerical benchmarks across the SK, 2D EA, and 3D EA models. Our evaluation focuses on the accuracy of physical quantities, ground-state optimization performance, and computational efficiency, with particular attention to disorder-dependent fluctuations and comparisons against established numerical methods.

\textbf{Energy estimates and numerical references.}
Energy accuracy and ground-state optimization are evaluated using the final sample-mean energy per spin and the minimum observed energy, respectively. Disorder-averaged energy estimates are reported with standard errors across instances. Where exact ground-state energies are available, we measure deviations from the reference values, including the geometric mean of absolute energy-density errors for 2D EA. Comparisons with IAMP~\cite{responseMontanariSK,responseIAMPImplementation2020} use matched disorder instances and identical Hamiltonian normalization, with energies reevaluated on discrete configurations in double precision.

\textbf{Overlap measurements.}
Beyond energy estimates, overlap distributions provide a more stringent test of whether FlashVAN captures the equilibrium statistical properties of spin-glass systems. The overlap $q$, fixed-disorder distribution $P_J(q)$, and disorder-averaged distribution $P(q)$ follow the definitions in Eq.~\ref{eq:overlap_distribution}. Overlaps are computed from independent configurations sharing the same disorder realization and temperature. As demonstrated in Results, FlashVAN reproduces the overlap statistics of established Monte Carlo benchmarks, including those for three-dimensional spin glasses. Reference-data sources and overlap-evaluation settings are summarized in Supplementary Table~10.

\textbf{Disorder averages and statistical errors.}
We further examine the small-overlap weight $I_L(T)$, defined in Eq.~\ref{eq:small_overlap_weight}, to characterize low-overlap configurations across disorder realizations. Configurations are grouped into disjoint replica pairs, and $\Pr_J(q^2<0.01)$ is estimated for each disorder instance before averaging across instances. Statistical uncertainties are estimated by bootstrap resampling over disorder instances rather than individual replica pairs. The resulting comparisons assess the consistency of FlashVAN with reference overlap statistics while accounting for disorder-dependent fluctuations.

\textbf{Optimization success and timing.}
Computational efficiency is assessed through the success probability and the time required to reach a prescribed threshold. A run succeeds when its best observed energy density matches the reference to five decimal places. Repeated runs determine the success probability, and $t_{0.9}$ denotes the mean runtime at the first tested budget reaching at least 90\% success. FlashVAN timings include all applicable training stages, following the reference timing convention~\cite{del2026demonstrating}. GA timings include calibrated published results and direct RTX~5090 measurements. The MADE transfer test retains the population-assisted training strategy while replacing the Transformer backbone with MADE, using the same success criterion and timing convention (Supplementary Section~IV.B.1, Fig.~11 and Table~12). Per-instance optimization costs and mean overlap-training times are reported in Supplementary Tables~11 and~13.

\textbf{Data availability:}
The authors declare that the data supporting this study are available within the paper.

\textbf{Code availability:}
A PyTorch implementation of the present algorithm can be found %
at the GitHub repository, which will be publicly available upon the acceptance of the manuscript. 

\subsection*{Acknowledgments}
\addtocontents{toc}{\protect\setcounter{tocdepth}{0}}
\addtocontents{toc}{\protect\setcounter{tocdepth}{3}}
We thank Lei Wang, Yuliang Jin, Luciano Hugo Miranda Filho for their helpful communication. This work is supported by Project 12322501, 12575035 of the National Natural Science Foundation of China, and 2026NSFSCZY0124 of the Natural Science Foundation of Sichuan Province. 
The computational work is supported by the Center for HPC, University of Electronic Science and Technology of China. 

\subsection*{Author contributions}
\addtocontents{toc}{\protect\setcounter{tocdepth}{0}}
\addtocontents{toc}{\protect\setcounter{tocdepth}{3}}
Y.T., P.Z., J.L. had the original idea for this work. L.Z., W.D. and J.L. performed the study, and all authors contributed to the preparation of the manuscript.

\subsection*{Competing interests}
\addtocontents{toc}{\protect\setcounter{tocdepth}{0}}
\addtocontents{toc}{\protect\setcounter{tocdepth}{3}}

The authors declare no competing interests.

\subsection*{Additional information}
\addtocontents{toc}{\protect\setcounter{tocdepth}{0}}
\addtocontents{toc}{\protect\setcounter{tocdepth}{3}}

\textbf{Supplementary information} The online version
contains supplementary material available at [URL will be inserted by publisher].

\textbf{Correspondence and requests for materials} should be addressed to Ying Tang.

\textbf{Reprints and permission information}  is available online at  [URL will be inserted by publisher].

\bibliography{FlashVAN}

\end{document}


\title{Supplementary Information: Scalable Physics-Inspired Transformers for Spin Glasses}

\author{Lu Zhong}
\altaffiliation{These authors contributed equally}
\affiliation{Institute of Fundamental and Frontier Sciences, University of Electronic Science and Technology of China, Chengdu 611731, China}
\author{Wenli Duan}
\altaffiliation{These authors contributed equally}
\affiliation{Institute of Fundamental and Frontier Sciences, University of Electronic Science and Technology of China, Chengdu 611731, China}
\author{Jing Liu}
\email[Corresponding authors: ]{jing.liu@bupt.edu.cn}
\affiliation{School of Physical Science and Technology, Beijing University of Posts and Telecommunications, Beijing 102206, China}
\affiliation{Institute of Theoretical Physics, Chinese Academy of Sciences, Beijing 100190, China}

\author{Pan Zhang}
\email[Corresponding authors: ]{panzhang@itp.ac.cn}
\affiliation{Institute of Theoretical Physics, Chinese Academy of Sciences, Beijing 100190, China}
\affiliation{School of Fundamental Physics and Mathematical Sciences, Hangzhou Institute for Advanced Study, UCAS, Hangzhou 310024, China}

\author{Ying Tang}
\email[Corresponding authors: ]{jamestang23@gmail.com}
\affiliation{Institute of Fundamental and Frontier Sciences, University of Electronic Science and Technology of China, Chengdu 611731, China}
\affiliation{School of Physics, University of Electronic Science and Technology of China, Chengdu 611731, China}
\affiliation{Key Laboratory of Quantum Physics and Photonic Quantum Information, Ministry of Education, University of Electronic Science and Technology of China, Chengdu 611731, China}
\affiliation{Non-classical Information Science Basic Discipline Research Center of Sichuan Province, University of Electronic Science and Technology of China, Chengdu 611731, China}

\maketitle
\tableofcontents

\clearpage

This Supplementary Information describes the architectural components and controlled comparisons in Section~\ref{sec:Model_details}, the training settings for each physical task in Section~\ref{sec:si-training}, additional physical benchmarks in Section~\ref{sec:si-physical}, and computational performance and neural baselines in Section~\ref{sec:hyparameters_details}. Model and training parameters are specified by task rather than through a universal parameter set.

\section{Model architecture and positional representations}
\label{sec:Model_details}
\subsection{Attention computation and interaction-based masks}
\label{sec:si-attention}

\label{sec:revised-attention}
Given a spin configuration $\boldsymbol{\sigma}=\{\sigma_1,\ldots,\sigma_N\}$, spin or patch states are mapped to input vectors through concatenated state and positional embeddings. For lattice models, spins or patches are ordered along a raster scan. Let $n$ be the resulting sequence length, $n_h$ the number of attention heads, $d_h$ the dimension per head, and $\mathbf h_t$ the input to the attention block for token $t$. Query, key, and value projections are
\begin{equation}
\mathbf a_t=\mathbf W_A\mathbf h_t=[\mathbf a_{t,1};\ldots;\mathbf a_{t,n_h}],\qquad A\in\{Q,K,V\},\quad\mathbf a\in\{\mathbf q,\mathbf k,\mathbf v\}.
\end{equation}
The matrices $\mathbf W_A\in\mathbb R^{n_hd_h\times d}$ project the input representation to the corresponding head features, and $[\cdot;\cdot]$ denotes concatenation. With $\mathcal A(t)$ the preceding tokens retained by the causal mask, the attention outputs are
\begin{equation}
\begin{aligned}
\mathbf o_{t,i}&=\sum_{j\in\mathcal A(t)}\operatorname{Softmax}_{j\in\mathcal A(t)}\!\left(\frac{\mathbf q_{t,i}^{\mathsf T}\mathbf k_{j,i}}{\sqrt{d_h}}\right)\mathbf v_{j,i},\\
\mathbf u_t&=[\mathbf o_{t,1};\ldots;\mathbf o_{t,n_h}].
\end{aligned}
\end{equation}
Full causal attention uses $\mathcal A(t)=\{1,\ldots,t\}$; a boundary-based sparse mask restricts this set according to the conditioning structure discussed in the main text. The final output projection $\mathbf W_O$ is bypassed in FlashVAN.

For nearest-neighbor lattice models, the ordered Markov boundary discussed in the main-text Methods identifies the sampled spins that screen the remaining past from the next spin. Under raster-scan ordering and open boundary conditions, its size scales as $\mathcal O(L)$ in 2D and $\mathcal O(L^2)$ in 3D. This motivates an attention mask retaining the corresponding boundary region. For a sliding-window implementation of width $W$, the attention weights obey
\begin{equation}
A_{ij}=0\quad\text{when}\quad i-j>W.
\end{equation}
The window must cover the required boundary in the actual spin or patch ordering. Periodic bonds can connect distant sequence positions, so the open-boundary argument alone does not justify a fixed local window for periodic lattices. Fully connected SK interactions require full causal attention. For a sequence of $n$ tokens and window width $W$, a sliding-window mask reduces the attention work from $\mathcal O(n^2)$ to $\mathcal O(nW)$.

\begin{figure}[!htbp]
	\begin{center}
		\includegraphics[width=\columnwidth,height=0.58\textheight,keepaspectratio]{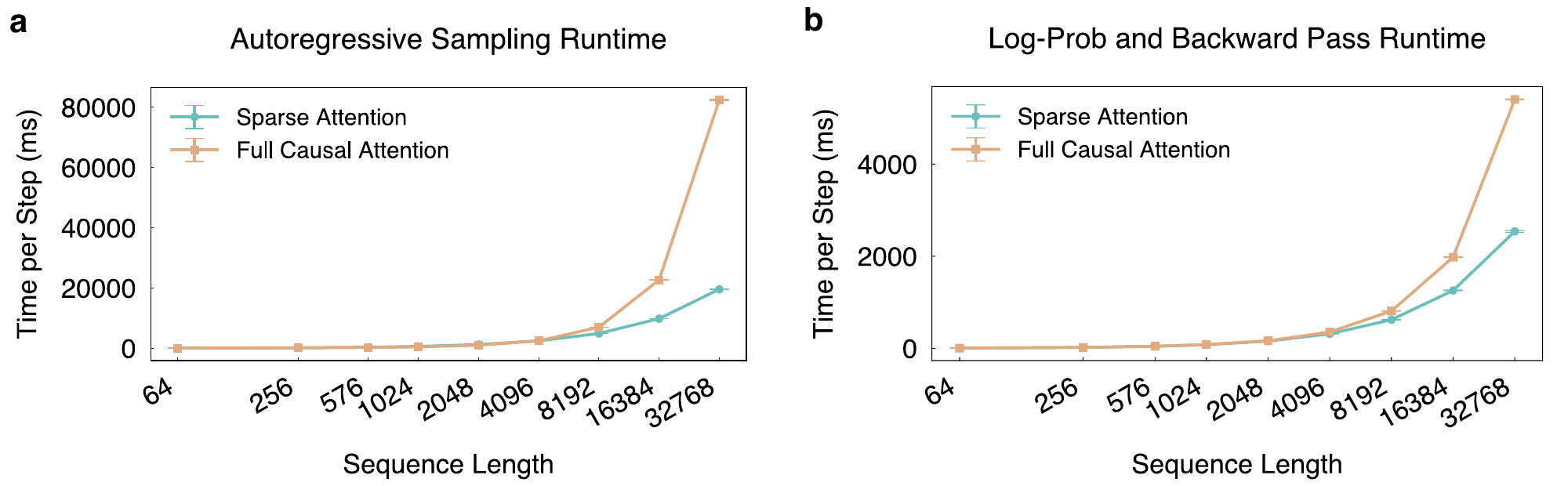}
	\end{center}
	\caption{\textbf{Runtime comparison between physics-inspired sparse attention and full causal attention.} (a) Autoregressive sampling runtime and (b) log-probability evaluation and backward pass runtime as a function of sequence length, for sparse attention and full causal attention.}
	\label{fig1A}
\end{figure}

\begin{figure}[!htbp]
	\begin{center}
		\includegraphics[width=\columnwidth,height=0.58\textheight,keepaspectratio]{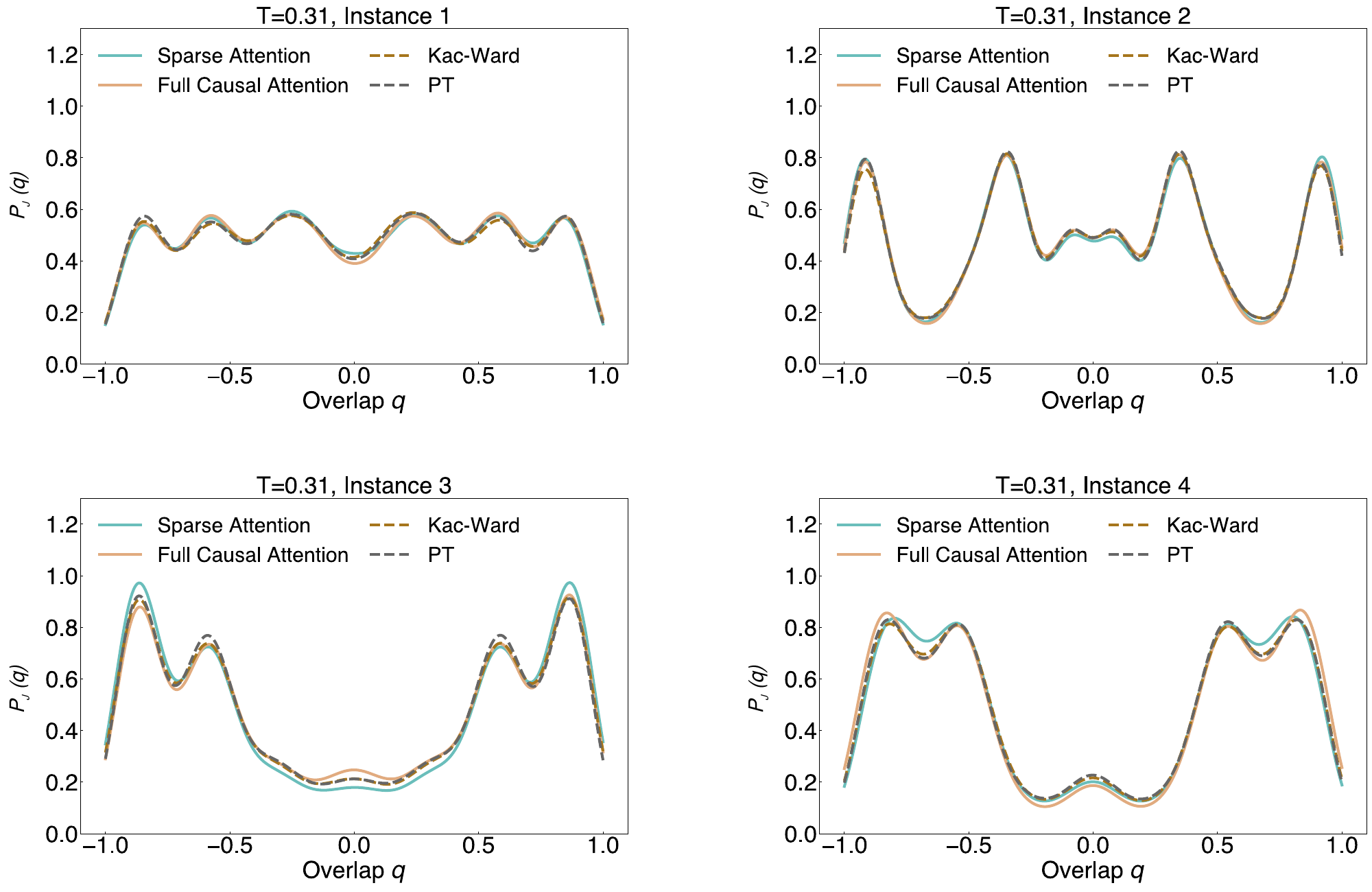}
	\end{center}
	\caption{\textbf{Effect of sparse attention on the overlap distribution $P_J(q)$.} Comparison between sparse attention and full causal attention in FlashVAN for four fixed disorder instances from Del Bono et al.~\cite{del2025nearest} at temperature $T = 0.31$. Both variants accurately reproduce the multi-peaked structure of $P_J(q)$ and show close agreement with the Kac-Ward-based sampler~\cite{responseKacWardSampler2026,PhysRev.88.1332} and parallel tempering (PT)~\cite{responsePT1996}.}
	\label{fig1A_2}
\end{figure}

Figure~\ref{fig1A} compares sparse and full attention runtimes in the 1D Ising benchmark. Figure~\ref{fig1A_2} compares their fixed-disorder overlap distributions in 2D EA. These controlled architecture comparisons assess the cost and observable accuracy of the mask choice; they are separate from the task-specific training protocols listed in Section~\ref{sec:si-training}.

\FloatBarrier

\subsection{Concatenated positional embeddings: controlled comparisons}\label{sec:si-pe}

To evaluate positional encodings, we compare concatenated positional embeddings (concatenated PE), which FlashVAN uses by default, with additive positional embeddings (Add PE) in three controlled experiments; the experimental traces appear in Supplementary Figures~\ref{fig1B_1}-\ref{fig1B_3}. All comparisons keep the model architecture and training hyperparameters identical; only the form of the positional encoding or the token/position embedding split is varied. When the total embedding dimension $d_{\rm model}$ is fixed, the allocation between token and positional components strongly affects performance at large system sizes, as illustrated in Supplementary Fig.~\ref{fig1B_1}. For small systems (for example $N=30$, $d_{\rm model}=32$), free-energy convergence is largely insensitive to the position-to-token ratio; by contrast, for large systems (for example $N=256$, $d_{\rm model}=256$), configurations that allocate excessive capacity to token embeddings exhibit degraded convergence and higher final free energy. Given the binary nature of the spin variables, a minimal token embedding ($d_{\rm token}=2$) suffices in practice (in the absence of patching) and frees capacity for positional encoding as system size grows.

\begin{figure}[!htbp]
	\begin{center}
		\includegraphics[width=\columnwidth,height=0.58\textheight,keepaspectratio]{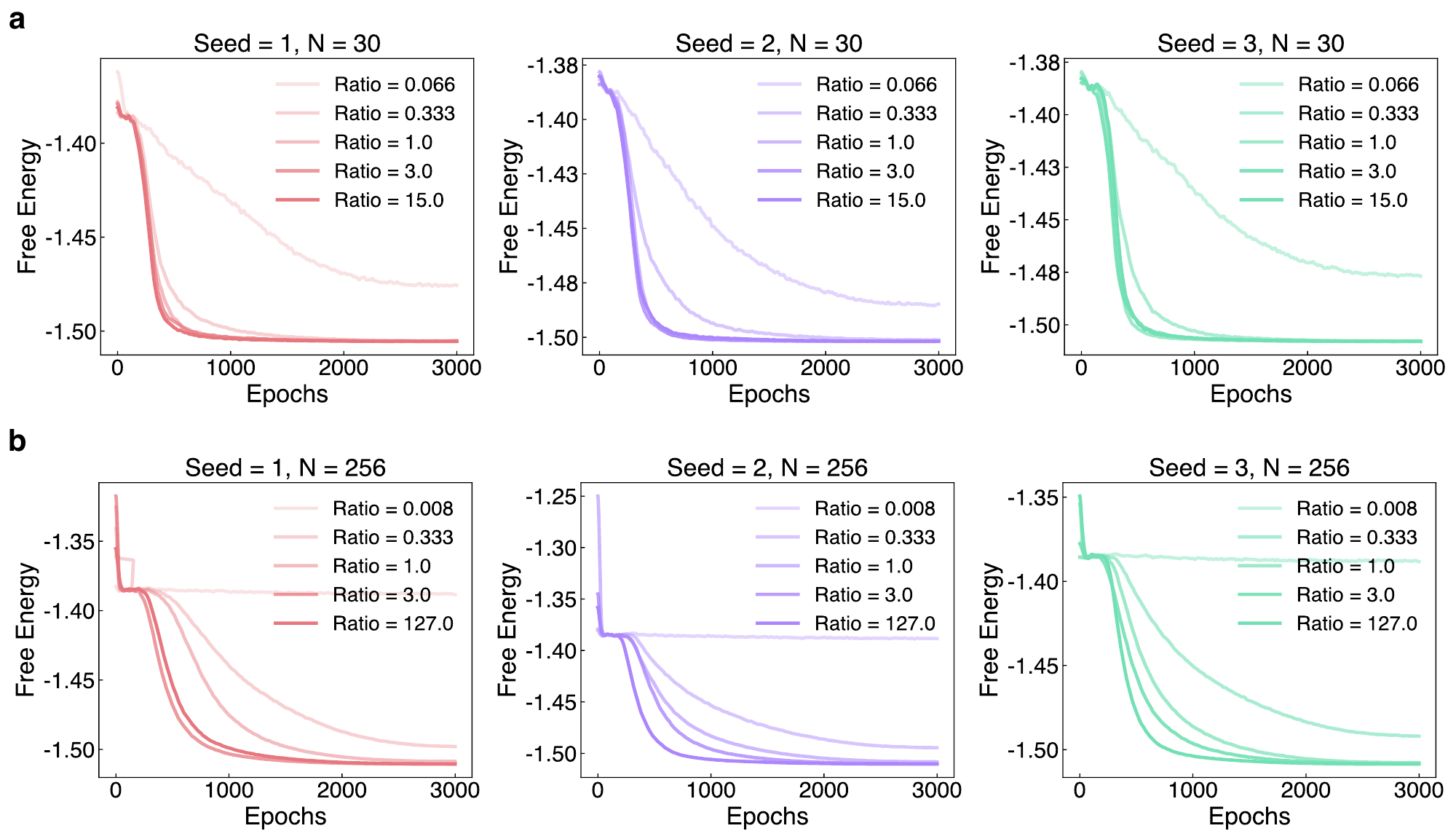}
	\end{center}
	\caption{\textbf{Effect of the positional-to-token embedding ratio on free energy convergence.} The total embedding dimension $d_{\rm model}$ is held constant, and Ratio is defined as $d_{\rm pos} / d_{\rm token}$. (a) Free-energy convergence trajectories for the SK model at system size $N=30$ under different Ratio values. (b) Corresponding results for $N = 256$. Columns correspond to different random seeds.}
	\label{fig1B_1}
\end{figure}

\FloatBarrier

In addition to allocation effects, concatenated PE yields more stable training trajectories and modestly faster convergence than Add PE, with this advantage most pronounced in low-temperature regimes where Add PE can exhibit oscillatory behavior (Supplementary Fig.~\ref{fig1B_2}).

\begin{figure}[!htbp]
	\begin{center}
		\includegraphics[width=\columnwidth,height=0.58\textheight,keepaspectratio]{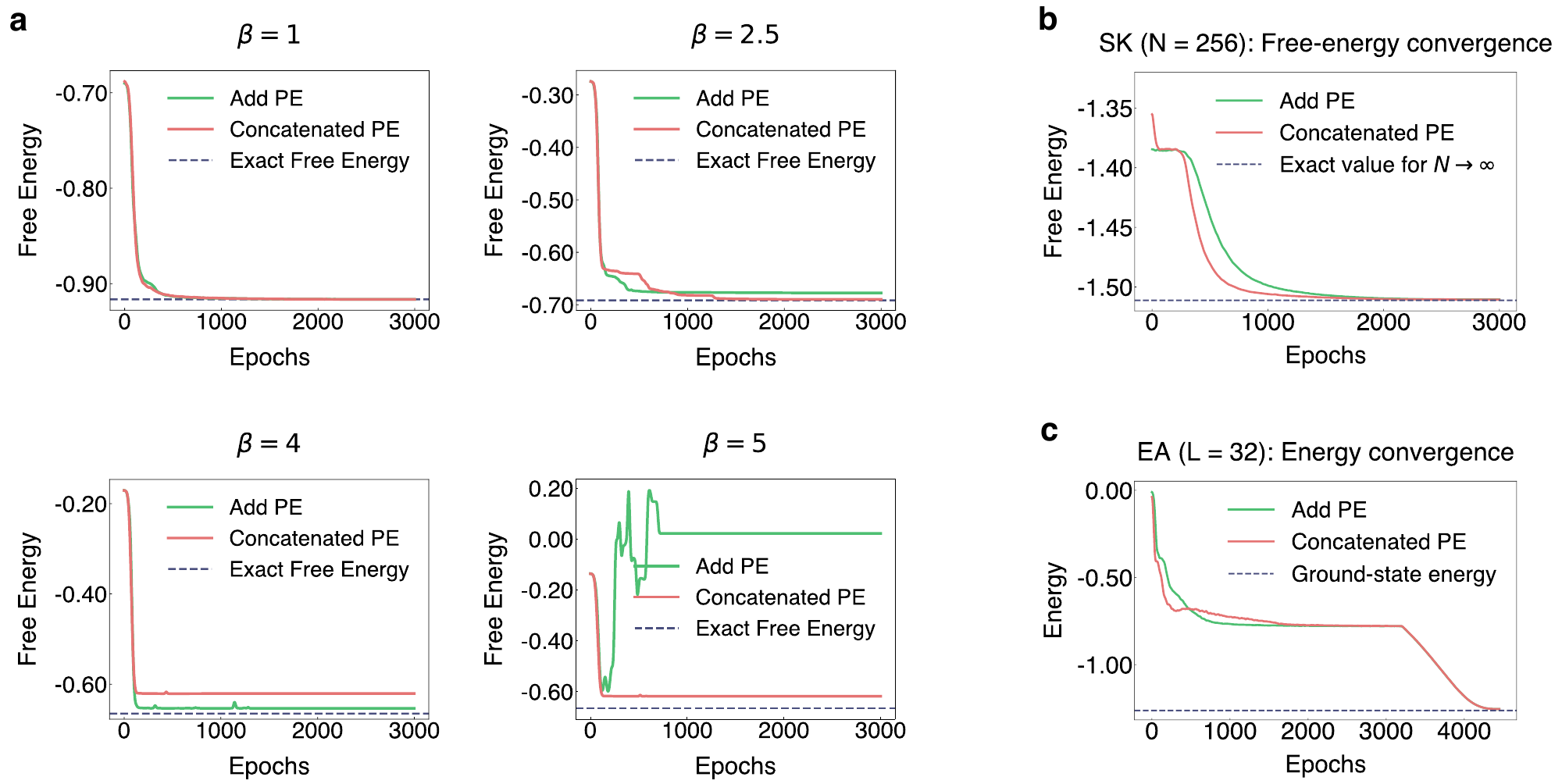}
	\end{center}
	\caption{\textbf{Comparison of two absolute positional-embedding (PE) schemes.} Green and red denote Add PE and concatenated PE, respectively; dashed lines indicate reference energies. (a) Free-energy convergence trajectories for the Sherrington-Kirkpatrick (SK) model at $N = 20$ under a fixed random seed, shown for inverse temperatures $\beta = 1, 2.5, 4, 5$. (b) Free-energy convergence for the SK model at $N = 256$ at $\beta = 0.5$. (c) Energy convergence during ground-state search in the 2D EA model.}
	\label{fig1B_2}
\end{figure}

\FloatBarrier

For the 2D EA overlap comparison in Supplementary Fig.~\ref{fig1B_3}, we use the fixed disorder instances provided by Del Bono et al.~\cite{del2025nearest}. The fixed-disorder overlap distributions $P_J(q)$ show that, when training converges, both embedding schemes recover the multipeaked structure and agree with the Kac-Ward-based sampler~\cite{responseKacWardSampler2026,PhysRev.88.1332} and parallel tempering (PT)~\cite{responsePT1996} references.

\begin{figure}[!htbp]
	\begin{center}
		\includegraphics[width=\columnwidth,height=0.58\textheight,keepaspectratio]{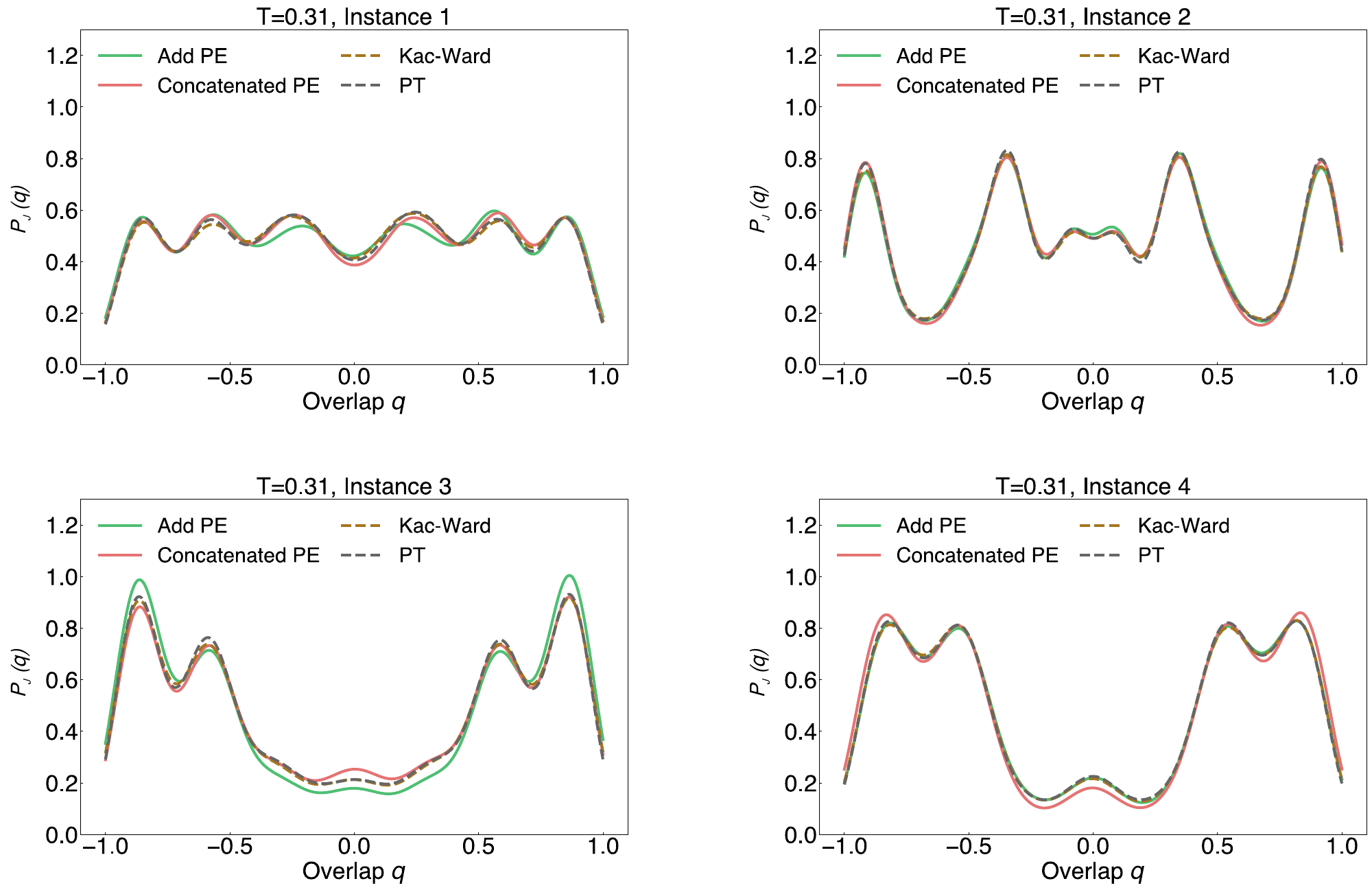}
	\end{center}
	\caption{\textbf{Comparison of Add and concatenated positional embeddings via fixed-disorder overlap distributions $P_J(q)$.} Curves show $P_J(q)$ for four disorder instances provided by Del Bono et al.~\cite{del2025nearest} at temperature $T = 0.31$. The Kac-Ward-based sampler~\cite{responseKacWardSampler2026,PhysRev.88.1332} and PT~\cite{responsePT1996} are included as references.}
	\label{fig1B_3}
\end{figure}

\FloatBarrier

Based on these empirical findings, we recommend allocating minimal dimensionality to token embeddings for binary spins (e.g. $d_{\rm token}=2$) and using concatenated PE for improved stability and scalability to large system sizes.

\subsection{Patching and spatial rotary positional embeddings}
\label{sec:revised-rope}
Rotary positional embedding (RoPE)~\cite{RoPE} introduces relative positional information by rotating queries and keys within attention. Here the coordinates refer to patches on the coarse lattice: a $2\times2$ patch produces a grid of size $(L/2)^2$, and a $2\times2\times2$ patch produces a grid of size $(L/2)^3$. Spin-state and absolute-position vectors are still concatenated at the model input. The SK configuration uses single-spin tokens without RoPE or patching.

Let $\mathbf q_{(m_i,n_i)}$ and $\mathbf k_{(m_j,n_j)}$ denote query and key vectors at two patch positions. In a four-dimensional example, the two spatial axes act on separate feature pairs through
\begin{equation}\label{eq:si-rope2d}
\mathbf R(m,n)=
\begin{bmatrix}
\cos(m\omega_x)&-\sin(m\omega_x)&0&0\\
\sin(m\omega_x)&\cos(m\omega_x)&0&0\\
0&0&\cos(n\omega_y)&-\sin(n\omega_y)\\
0&0&\sin(n\omega_y)&\cos(n\omega_y)
\end{bmatrix}.
\end{equation}
The rotated vectors $\widetilde{\mathbf q}_i=\mathbf R(m_i,n_i)\mathbf q_i$ and $\widetilde{\mathbf k}_j=\mathbf R(m_j,n_j)\mathbf k_j$ satisfy
\begin{equation}
\widetilde{\mathbf q}_i^{\mathsf T}\widetilde{\mathbf k}_j=\mathbf q_i^{\mathsf T}\mathbf R(m_j-m_i,n_j-n_i)\mathbf k_j.
\end{equation}
Thus the attention dot product depends on the relative coordinates. Higher-dimensional vectors are divided into even-dimensional feature groups for each spatial axis, with independent two-dimensional rotations within each group. In 3D, an additional group encodes the third axis.

For head dimension $d_h$, the 2D allocation is $d_x=d_h-2\lfloor d_h/4\rfloor$, $d_y=2\lfloor d_h/4\rfloor$. In 3D, $d_x=d_h-4\lfloor d_h/6\rfloor$ and $d_y=d_z=2\lfloor d_h/6\rfloor$. For an axis group of dimension $d_a$, the $j$th feature pair is rotated by $r_a\omega_{a,j}$, where
\begin{equation}
\omega_{a,j}=\theta^{-2j/d_a},\qquad j=0,\ldots,d_a/2-1,\qquad\theta=100000,
\end{equation}
and $r_a$ is the integer patch coordinate. These rotations act on queries and keys; values are unchanged.

Architectural symbols are defined in Table~\ref{tab:symbols}; their task-specific values are listed in Section~\ref{sec:si-training}.

\clearpage
\section{Population-assisted training protocols}
\label{sec:si-training}

\subsection{Notation for architecture and training settings}
Table~\ref{tab:symbols} defines the symbols and entries used in the task-specific configuration tables. A training update refers to one model-parameter update; MCS counts describe local Monte Carlo evolution of each pool configuration.

\begin{table}[!htbp]
\centering\small
\setlength{\tabcolsep}{4pt}
\renewcommand{\arraystretch}{1.10}
\caption{Definitions of architecture and training entries.}
\label{tab:symbols}
\fitsitable{%
\begin{tabular}{ll}
\toprule
Symbol / entry & Meaning\\
\midrule
$N$, $L$ & Number of spins and lattice linear size\\
Couplings & Distribution or source of the interaction strengths\\
Boundary conditions & Open (OBC) or periodic (PBC) lattice boundaries\\
No. of instances & Number of distinct disorder realizations\\
$n_{\rm layers}$ & Number of Transformer decoder layers\\
$n_{\rm heads}$ & Number of attention heads per layer\\
$d_{\rm token}$ & Spin-state or patch-state embedding dimension\\
$d_{\rm pos}$ & Absolute positional embedding dimension\\
$d_{\rm model}$ & Total input dimension, $d_{\rm token}+d_{\rm pos}$\\
Patch size & Number and arrangement of spins grouped into one token\\
RoPE & Rotary positional embedding within attention\\
Physics-inspired attention & Whether the interaction-based sparse mask is used\\
$\eta$ & Optimizer learning rate\\
$N_{\rm batch}$ & Number of configurations per training batch\\
\midrule
$T_0$ & Initial annealing temperature\\
$T_t$ & Target physical temperature\\
MCS & One attempted single-spin update per spin per configuration\\
Initial population & Number of configurations in the initial population\\
Initial MCS & Local Monte Carlo sweeps for initial population preparation\\
MLE pretraining updates & Number of maximum-likelihood initialization updates\\
Persistent pool size & Number of retained configurations in the training pool\\
Annealing schedule & Sequence of physical temperatures during training\\
Annealing updates & Number of model updates during temperature annealing\\
Injected samples per update & Number of model configurations replacing pool entries\\
Annealing MCS & Local Monte Carlo sweeps per annealing update\\
Annealing $w$ & Weight of the additional likelihood term during annealing\\
Refinement MCS & Local Monte Carlo sweeps per refinement update\\
Refinement $w$ & Likelihood weight during refinement\\
Refinement min / max & Minimum and maximum refinement update counts\\
Refinement probe interval & Updates between successive diagnostic evaluations\\
Distillation updates & Maximum-likelihood updates on the frozen refined pool\\
Final model configurations & Number of configurations generated for final evaluation\\
\midrule
$n_{\rm warmup}$ & High-temperature training updates in energy-scaling runs\\
$n_{\rm anneal}$ & Annealing levels or updates in the specified schedule\\
$n_{\rm eq}$ & Repeated updates per fixed-temperature annealing level\\
\bottomrule
\end{tabular}}
\end{table}

\subsection{Training stages and local updates}
\label{sec:si-training-stages}
Population-assisted autoregressive training begins with an initialization stage (Stage~0), followed by the three main stages: annealing, refinement, and distillation. Initialization prepares both the model and the persistent sample pool. The subsequent stages establish the feedback between model generation and population refinement, then transfer the refined population statistics into the directly sampleable model.

\textbf{Stage 0: Initialization.} We first draw an initial population of 131072 random spin configurations and relax them at the high temperature $T_0$ for 200 Monte Carlo sweeps. Standard local Metropolis Monte Carlo (LMC) updates~\cite{Metropolis1953} accept a proposed single-spin flip with probability $\min\{1,\exp(-\Delta E/T)\}$, and one sweep (MCS) is one attempted update per spin per configuration. The EA implementations update the two checkerboard sublattices in turn; SK updates spins sequentially with refreshed local fields. FlashVAN is then pretrained for 1000 maximum-likelihood updates, using minibatches drawn from this high-temperature population.

After pretraining, we select 16384 configurations uniformly at random without replacement from the initial population to initialize the persistent pool. Thus the larger initial population supplies data for pretraining, while only the selected subset is propagated through the subsequent training stages. The persistent pool remains at this smaller size during annealing and refinement, with its configurations evolving and being replaced as described below. The entries \textit{Initial population} and \textit{Persistent pool size} in the task tables refer to these two different roles.

\textbf{Stage 1: Annealing.} The temperature decreases from $T_0$ to the target $T_t$. At a change from inverse temperature $\beta_{k-1}$ to $\beta_k$, pool configurations are reweighted by $\exp[-(\beta_k-\beta_{k-1})E]$ and resampled at fixed pool size. Fresh autoregressive samples from the current model replace a subset of pool entries, and LMC updates refine the resulting population at the new temperature. Batches from this updated pool train the model using the population-assisted objective defined in the main-text Methods. The model therefore contributes new configurations to the pool and learns from the configurations explored through local relaxation. The injection counts, LMC sweeps, likelihood weights, and annealing budgets depend on the task and are specified in the corresponding tables.

\textbf{Stage 2: Refinement.} At $T_t$, model-sample injection and population resampling are suspended. The existing pool continues to evolve through LMC, and model training continues on batches from this pool using the same population-assisted objective. This stage allows further local relaxation without replacing pool configurations with fresh model samples. Refinement stops dynamically according to the diagnostics and stopping rules in Section~\ref{sec:si-refinement}. We also impose minimum and maximum numbers of refinement updates, listed as \textit{Refinement min / max} in the task tables: the stopping criterion is applied only after the minimum duration, and refinement ends at the maximum if it has not stopped earlier.

\textbf{Stage 3: Distillation.} The refined pool is frozen, and the same autoregressive model is trained by pure maximum likelihood on configurations drawn from this pool. The temperature remains fixed, with no further LMC updates or model-sample injection. This final fit transfers the pool statistics into the generative model. Finite-temperature observables are subsequently evaluated using configurations sampled directly from the trained model.

The task tables specify which stages are used for each calculation. Ground-state optimization uses task-specific stage choices; these choices do not change the distinction between the initial population and the persistent training pool.

\subsection{Refinement diagnostics and stopping rules}
\label{sec:si-refinement}
For independent Gaussian bonds of mean $\mu$ and variance $\sigma_J^2$, Gaussian integration by parts gives the disorder-averaged equilibrium relation
\begin{equation}
\mathbb E_J\!\left[\frac{\langle E\rangle_J}{B}+\mu\langle m_b\rangle_J+
\frac{\sigma_J^2}{T}(1-\langle q_\ell\rangle_J)\right]=0,
\end{equation}
where $B$ is the number of bonds, $m_b=B^{-1}\sum_{\langle ij\rangle}\sigma_i\sigma_j$, and $q_\ell=B^{-1}\sum_{\langle ij\rangle}\sigma_i^1\sigma_j^1\sigma_i^2\sigma_j^2$. The centered unit-variance cubic-lattice form is $S=e_{\rm pool}/3+(1-q_{\ell,\rm pool})/T$ for periodic boundaries~\cite{chilin2026true}. The normalized Gaussian SK counterpart is $S=e_{\rm pool}+(1-\langle q^2\rangle_{\rm pool})/(2T)$. For the supplied 2D overlap instances, the pool diagnostic retains the bond-mean and variance terms.

Finite-temperature refinement monitors the change $|S(t)-S(t/4)|$ against a tolerance of $0.003$ at two consecutive probes. The probe intervals and duration bounds are specified in the task tables. Here $t$ counts refinement updates, and $S$ is estimated from the pool. Stagnation is an operational stopping rule: the equilibrium identity concerns a disorder average and does not certify equilibrium for an individual realization. Ground-state protocols instead monitor stagnation of the best pool energy: the SK optimization runs require 100 updates without improvement, and the 3D search-cost runs require 50. Distillation uses no LMC updates or model-sample injection.

\clearpage
\subsection{SK configurations and training protocols}
Table~\ref{tab:si-sk-protocols} separates the finite-temperature overlap protocol from the two SK optimization benchmarks. Table~\ref{tab:SK settings} gives the size-dependent configurations and timings for the bimodal energy comparison.

\begin{table}[!htbp]
\centering\small
\setlength{\tabcolsep}{4pt}
\renewcommand{\arraystretch}{1.15}
\caption{SK architecture and training settings by task.}
\label{tab:si-sk-protocols}
\fitsitable{%
\begin{tabular}{lccc}
\toprule
Parameter & Overlap sampling & Bimodal energy scaling & Gaussian optimization\\
\midrule
Result / measurement & Fig.~2b & Fig.~2c & IAMP comparison\\
System size & $N=256$ & $N=63,127,255$ & $N=256$\\
Couplings & Gaussian & $\pm1$ & Gaussian\\
Boundary conditions & Not applicable & Not applicable & Not applicable\\
No. of instances & 10 & 200, 100, 50 & 30\\
$n_{\rm layers}$ / $n_{\rm heads}$ & 2 / 2 & 2 / 2 & 2 / 2\\
$d_{\rm token}$ / $d_{\rm pos}$ & 2 / 254 & Table~\ref{tab:SK settings} & 2 / 254\\
$d_{\rm model}$ & 256 & Table~\ref{tab:SK settings} & 256\\
Patch size / RoPE & Single spin / None & Single spin / None & Single spin / None\\
Physics-inspired attention & No (full causal) & No (full causal) & No (full causal)\\
Optimizer / learning rate & Muon / $10^{-3}$ & Muon / $10^{-3}$ & Muon / $10^{-3}$\\
Batch size & 2048 & 2048 & 2048\\
\midrule
Initial temperature $T_0$ & 10 & 10 & 10\\
Initial population & 131072 & 131072 & 131072\\
Initial MCS & 200 & 200 & 200\\
MLE pretraining updates & 1000 & 1000 & 1000\\
Persistent pool size & 16384 & 16384 & 16384\\
Target temperature $T_t$ & $2,1,2/3,1/2$ & $0.01$ & $0.01$\\
Annealing schedule & Logarithmic & Logarithmic & Logarithmic\\
Annealing updates & 2000 & 2000 & 2000\\
Injected samples per update & \makecell{256 at $T=2$;\\2048 otherwise} & 2048 & 2048\\
Annealing MCS & 30 & 30 & 30\\
Annealing $w$ & 0.9 & 0.9 & 0.9\\
Refinement MCS & 120 & 120 & 120\\
Refinement $w$ & 0.9 & 0.9 & 0.9\\
Refinement min / max & 400 / 2000 & 100 / 2000 & 100 / 2000\\
Refinement probe interval & 25 updates & Not used & Not used\\
Distillation updates & 1500 & 0 & 0\\
Final model configurations & 20000 & 20000 & 20000\\
\bottomrule
\end{tabular}}
\par\smallskip
\raggedright\footnotesize Figure~2c reports final sample-mean energies; the matched IAMP comparison evaluates best discrete configurations, as described in Section~\ref{sec:si-iamp}.
\end{table}

\begin{table}[!htbp]
\centering
\caption{Hyperparameters and training times for SK ground-state search at different system sizes $N$.}
\label{tab:SK settings}
\begin{tabular}{lccc}
\toprule
Parameter &
\makecell{$N=63$} &
\makecell{$N=127$} &
\makecell{$N=255$} \\
\midrule
$d_{\rm token}$ & 2 & 2 & 2 \\
$d_{\rm model}$ & 64 & 128 & 256 \\
$n_{\rm layers}$ & 2 & 2 & 2 \\
$d_{\rm pos}$ & 62 & 126 & 254 \\
$n_{\rm heads}$ & 2 & 2 & 2 \\
Annealing updates & 2000 & 2000 & 2000 \\
Wall-clock time per training step & \SI{0.081}{s} & \SI{0.201}{s} & \SI{0.640}{s} \\
Total wall-clock training time & \SI{3.212}{\minute} & \SI{7.786}{\minute} & \SI{25.530}{\minute} \\

\bottomrule
\end{tabular}
\end{table}

\clearpage
\subsection{2D EA configurations and training protocols}
Table~\ref{tab:si-2dea-protocols} distinguishes the overlap benchmark from the population-assisted energy-timing tests in Table~\ref{tab:2DEA settings}. These timing tests do not replace the full energy-scaling ensemble in main-text Fig.~3c, whose variational annealing configurations are listed separately in Table~\ref{tab:2dea-energy-scaling}.

\begin{table}[!htbp]
\centering\small
\setlength{\tabcolsep}{4pt}
\renewcommand{\arraystretch}{1.15}
\caption{2D EA overlap and population-assisted energy-timing configurations.}
\label{tab:si-2dea-protocols}
\fitsitable{%
\begin{tabular}{lcc}
\toprule
Parameter & Overlap sampling & Energy timing tests\\
\midrule
Result / measurement & Fig.~3b & Separate PBC timing runs\\
System size & $L=16$ & $L=8,16,24,32,48,60$\\
Couplings & Provided fixed instances & Gaussian\\
Boundary conditions & OBC & PBC\\
No. of instances & 4 & One per size\\
$n_{\rm layers}$ / $n_{\rm heads}$ & 2 / 4 & Table~\ref{tab:2DEA settings}\\
$d_{\rm token}$ / $d_{\rm pos}$ & 16 / 112 & Table~\ref{tab:2DEA settings}\\
$d_{\rm model}$ & 128 & Table~\ref{tab:2DEA settings}\\
Patch size / RoPE & $2\times2$ / 2D & $2\times2$ / 2D\\
Physics-inspired attention & Yes & Yes\\
Optimizer / learning rate & Muon / $10^{-3}$ & Muon / $10^{-3}$\\
Batch size & 1024 & Table~\ref{tab:2DEA settings}\\
\midrule
Initial temperature $T_0$ & 1.92 & 1.92\\
Initial population & 131072 & 131072\\
Initial MCS & 200 & 200\\
MLE pretraining updates & 1000 & 1000\\
Persistent pool size & 16384 & 16384\\
Target temperature $T_t$ & 0.3125 & 0.01\\
Annealing schedule & Logarithmic & Logarithmic\\
Annealing updates & 1000 & Table~\ref{tab:2DEA settings}\\
Injected samples per update & 2048 & 2048\\
Annealing MCS & 30 & 30\\
Annealing $w$ & 0.9 & 0.9\\
Refinement MCS & 120 & Omitted\\
Refinement $w$ & 0.9 & Not applicable\\
Refinement min / max & 400 / 2000 & Not applicable\\
Refinement probe interval & 25 updates & Not applicable\\
Distillation updates & 1500 & 0\\
Final model configurations & 200000 & 100000\\
\bottomrule
\end{tabular}}
\par\smallskip
\raggedright\footnotesize The stopping diagnostic for the fixed overlap instances includes the bond-mean and variance terms. Table~\ref{tab:2DEA settings} lists the separately measured PBC timing runs. These single-instance timing tests do not replace the full energy-scaling ensemble in main-text Fig.~3c.
\end{table}

\begin{table}[!htbp]
\centering
\caption{Size-dependent 2D EA energy configurations and PBC timing measurements at different system sizes $L$.}
\label{tab:2DEA settings}
\begin{tabular}{lcccccc}
\toprule
Parameter &
\makecell{$L=8$} &
\makecell{$L=16$} &
\makecell{$L=24$} &
\makecell{$L=32$} &
\makecell{$L=48$} &
\makecell{$L=60$} \\
\midrule
$d_{\rm model}$ & 64 & 128 & 256 & 320 & 160 & 192 \\
$n_{\rm layers}$ & 2 & 2 & 2 & 2 & 2 & 2 \\
$d_{\rm pos}$ & 62 & 126 & 254 & 318 & 144 & 176 \\
$d_{\rm token}$ & 2 & 2 & 2 & 2 & 16 & 16 \\
$n_{\rm heads}$ & 2 & 4 & 4 & 4 & 4 & 4 \\
MLE pretraining updates & 1000 & 1000 & 1000 & 1000 & 1000 & 1000 \\
Annealing updates & 500 & 600 & 1100 & 1200 & 2000 & 2000 \\
Updates per temperature & 1 & 1 & 1 & 1 & 1 & 1 \\
$N_{\rm batch}$ & 1024 & 1024 & 1024 & 1024 & 2048 & 2048 \\
Wall-clock time per training step & \SI{0.024}{s} & \SI{0.064}{s} & \SI{0.172}{s} & \SI{0.380}{s} & \SI{1.236}{s} & \SI{2.710}{s} \\
Total wall-clock training time & \SI{0.442}{\minute} & \SI{0.971}{\minute} & \SI{5.162}{\minute} & \SI{12.831}{\minute} & \SI{0.836}{\hour} & \SI{1.821}{\hour} \\

\bottomrule
\end{tabular}
\end{table}

\begin{table}[!htbp]
\centering
\caption{Configurations for the 2D EA energy-scaling dataset, with two decoder layers at every size.}
\label{tab:2dea-energy-scaling}
\begin{tabular}{lcccccc}
\toprule
Parameter &
\makecell{$L=8$} &
\makecell{$L=16$} &
\makecell{$L=24$} &
\makecell{$L=32$} &
\makecell{$L=48$} &
\makecell{$L=60$} \\
\midrule
$d_{\rm pos}$ & 62 & 126 & 254 & 318 & 144 & 176 \\
$d_{\rm token}$ & 2 & 2 & 2 & 2 & 16 & 16 \\
$n_{\rm heads}$ & 2 & 4 & 4 & 4 & 4 & 4\\
$n_{\rm warmup}$ & 2000 & 2000 & 3000 & 3200 & 4500 & 4500 \\
$n_{\rm anneal}$ & 100 & 120 & 220 & 240 & 2000 & 2000 \\
$n_{\rm eq}$ & 5 & 5 & 5 & 5 & 0 & 0 \\
$N_{\rm batch}$ & 1024 & 1024 & 1024 & 1024 & 2048 & 2048 \\

\bottomrule
\end{tabular}
\end{table}

\clearpage
\subsection{3D EA configurations and training protocols}
Table~\ref{tab:si-3dea-protocols} compares the population-assisted protocols used for overlap sampling, optimization costs, and small-overlap weights. The energy-scaling dataset in main-text Fig.~4c retains the separate variational annealing configurations in Table~\ref{tab:3DEA settings}.

\begin{table}[!htbp]
\centering\small
\setlength{\tabcolsep}{4pt}
\renewcommand{\arraystretch}{1.15}
\caption{Population-assisted 3D EA configurations for overlap sampling and optimization.}
\label{tab:si-3dea-protocols}
\fitsitable{%
\begin{tabular}{lccc}
\toprule
Parameter & Overlap sampling & \makecell{Ground-state\\search cost} & Small-overlap weight\\
\midrule
Result / measurement & Fig.~4b & Fig.~4d & Fig.~4e; size extension\\
System size & $L=10$ & $L=14$ & $L=14,16$\\
Couplings & Gaussian & Gaussian & Gaussian\\
Boundary conditions & PBC & PBC & PBC\\
No. of instances & 1 & 10 & \makecell{$L=14$: 60 or 10\textsuperscript{a}\\$L=16$: 10}\\
$n_{\rm layers}$ / $n_{\rm heads}$ & 2 / 2 & 2 / 2 & 2 / 2\\
$d_{\rm token}$ / $d_{\rm pos}$ & 64 / 192 & 64 / 192 & 64 / 192\\
$d_{\rm model}$ & 256 & 256 & 256\\
Patch size / RoPE & $2\times2\times2$ / 3D & $2\times2\times2$ / 3D & $2\times2\times2$ / 3D\\
Physics-inspired attention & Yes & Yes & Yes\\
Optimizer / learning rate & Muon / $10^{-3}$ & Muon / $10^{-3}$ & Muon / $10^{-3}$\\
Batch size & 2048 & 2048 & 2048\\
\midrule
Initial temperature $T_0$ & 1.92 & 1.92 & 1.92\\
Initial population & 131072 & 131072 & 131072\\
Initial MCS & 200 & 200 & 200\\
MLE pretraining updates & 1000 & 1000 & 1000\\
Persistent pool size & 16384 & 16384 & 16384\\
Target temperature $T_t$ & $1.92,0.77,0.29,0.10$ & 0.1 & Seven temperatures\textsuperscript{a}\\
Annealing schedule & Logarithmic & Logarithmic & Logarithmic\\
Annealing updates & \makecell{8000 at $T=0.77$;\\2000 otherwise} & \makecell{Instance-specific\\budget scan\textsuperscript{b}} & 2000\\
Injected samples per update & \makecell{256 at $T=1.92,0.77$;\\2048 at $T=0.29,0.10$} & 2048 & \makecell{2048 at $T\le0.95$;\\256 otherwise}\\
Annealing MCS & \makecell{5 at $T=0.77$;\\30 otherwise} & 30 & 30\\
Annealing $w$ & 0.9 & \makecell{0.2 at $T>0.2$;\\0.5 otherwise} & 0.9\\
Refinement MCS & 120 & 120 & 120\\
Refinement $w$ & 0.9 & 0.5 & 0.9\\
Refinement min / max & 400 / 2000 & 50 / 2000 & 400 / 2000\\
Refinement probe interval & 25 updates & 25 updates & 25 updates\\
Distillation updates & 1500 & 0 & 1500\\
Final model configurations & 20000 & 131072 & 20000\\
\bottomrule
\end{tabular}}
\par\smallskip
\raggedright\footnotesize
\textsuperscript{a} At $L=14$, 60 instances are used at each of $T=0.2915,0.3831,0.5051$, and ten at each of $T=0.7186,0.9627,1.1763,1.2983$. At $L=16$, ten instances are used at each of these seven temperatures.
\par\smallskip
\textsuperscript{b} The annealing budget is varied separately for each instance, with ten runs per tested budget. The attainment time uses the first budget reaching at least 90\% success.
\end{table}

\begin{table}[!htbp]
\centering
\caption{Configurations and per-step timings for the 3D EA energy-scaling dataset.}
\label{tab:3DEA settings}
\begin{tabular}{lcccccc}
\toprule
Parameter &
\makecell{$L=4$} &
\makecell{$L=6$} &
\makecell{$L=8$} &
\makecell{$L=10$} &
\makecell{$L=14$} &
\makecell{$L=16$} \\
\midrule
$d_{\rm pos}$ & 30 & 126 & 124 & 96 & 192 & 192 \\
$d_{\rm token}$ & 2 & 2 & 4 & 32 & 64 & 64 \\
$n_{\rm heads}$ & 2 & 2 & 2 & 2 & 2 & 2\\
$n_{\rm layers}$ & 3 & 3 & 3 & 2 & 2 & 2\\
$n_{\rm warmup}$ & 2000 &  2000 &  10000&  5000 & 5000 & 5000  \\
$n_{\rm anneal}$ & 100 &  100 &  300 &  2000 & 2000 & 2100   \\
$n_{\rm eq}$  & 5 &  5&  5& 1& 1 & 1  \\
$N_{\rm batch}$  & 1024 & 1024 & 2048 & 2048 & 2048 &  2048  \\
Wall-clock time per training step & \SI{0.028}{s} & \SI{0.052}{s} & \SI{0.132}{s} & \SI{0.179}{s} & \SI{0.819}{s} & \SI{1.290}{s} \\

\bottomrule
\end{tabular}
\end{table}

\subsection{Optimizer and learning-rate comparisons}
\label{sec:Optimizer_details}
\label{sec:Optimizer_comparison}
\begin{figure}[!htbp]
	\begin{center}
		\includegraphics[width=\columnwidth,height=0.58\textheight,keepaspectratio]{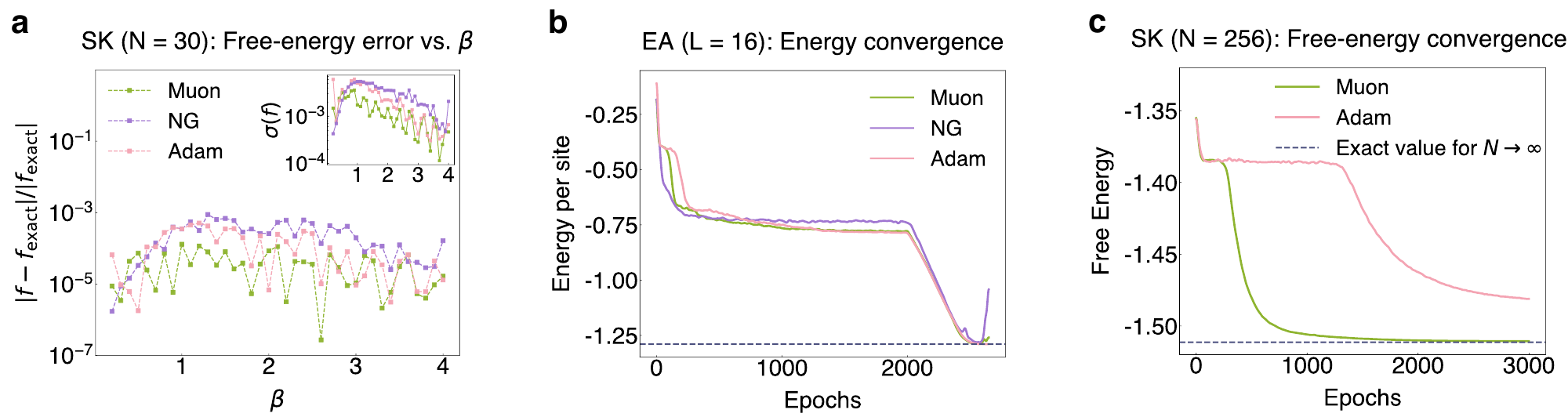}
	\end{center}
	\caption{\textbf{Effect of optimizer choice on training convergence.} Dashed lines indicate reference energies. (a) Relative free energy error $\left|f - f_{\mathrm{exact}}\right| / \left| f_{\mathrm{exact}}\right|$ for the SK model (N = 30) as a function of inverse temperature $\beta$; the inset shows the standard deviation $\sigma (f)$. Green, purple and pink curves denote the Muon, NG and Adam optimizers, respectively. (b) Energy convergence per site for the 2D EA model at $L = 16$. (c) Free-energy convergence for the SK model at $N = 256$. Under identical training and model configurations, NG could not be run at this scale due to GPU memory limitations. }
	\label{fig3A}
\end{figure}

\begin{figure}[!htbp]
	\begin{center}
		\includegraphics[width=\columnwidth,height=0.58\textheight,keepaspectratio]{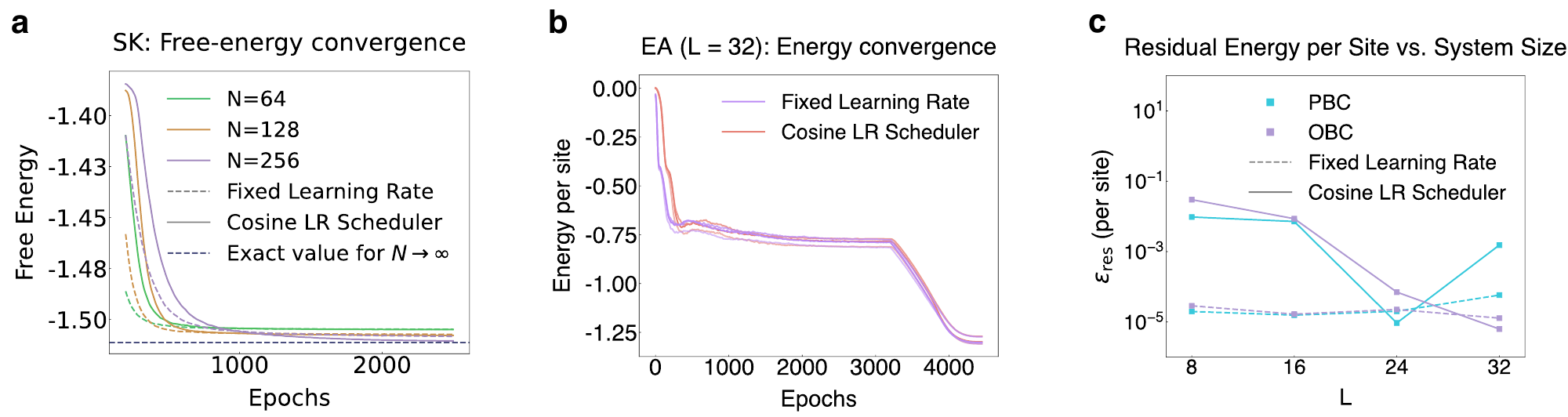}
	\end{center}
	\caption{\textbf{Impact of learning rate scheduler on training convergence and accuracy.} All panels compare a fixed learning rate of $1 \times 10^{-3}$ with a cosine learning rate scheduler using a warm-up phase of 300 steps. (a) Free-energy convergence for the SK model, showing training trajectories for $N=64,128,256$ with dashed horizontal lines marking reference free energies. (b) Energy convergence per site for the 2D EA model at $L = 32$. (c) Residual energy per site as a function of system size $L$, comparing PBC and OBC. }
	\label{fig3B}
\end{figure}

The controlled comparisons in Figs.~\ref{fig3A} and~\ref{fig3B} examine optimizer and learning-rate choices. They support the choice of numerical optimizer but do not define the task-specific population-assisted protocols above. Figure~\ref{fig3A} compares Adam~\cite{kingma2014adam}, improved natural-gradient descent~\cite{PhysRevE.111.025304}, and Muon~\cite{liu2025muon}. For the $N=30$ SK free-energy task, exact enumeration provides the reference, and the selected learning rates are $10^{-3}$, $10^{-2}$, and $10^{-3}$, respectively. For the $L=16$ 2D EA energy task, the corresponding rates are $10^{-3}$, $10^{-1}$, and $10^{-2}$. The $N=256$ SK comparison uses $10^{-3}$ for Adam and Muon; natural-gradient training exceeded the available GPU memory in that configuration.

The improved natural-gradient implementation uses a batch-space matrix rather than explicitly inverting a parameter-space matrix. Computing and storing per-sample gradients still imposes substantial memory costs. Muon offers a practical balance between convergence and memory consumption in these comparisons.

Figure~\ref{fig3B} compares a fixed learning rate of $10^{-3}$ with 300 warmup updates from $10^{-5}$ to $10^{-3}$ followed by cosine decay with $T_{\max}=2700$ and minimum rate $10^{-5}$. The schedule improves fixed-temperature SK convergence in the tested setting but can impair energy optimization during temperature annealing. These comparisons distinguish learning-rate scheduling from physical-temperature annealing; the settings for each main-text task are specified in the task tables.

\clearpage
\section{Additional physical benchmarks}
\label{sec:si-physical}
\subsection{SK overlap distributions and PT references}
\label{sec:si-sk-overlap}

The PT calculation uses the same Gaussian coupling matrices and Hamiltonian normalization as FlashVAN, following the exchange Monte Carlo method~\cite{responsePT1996}. The reference protocols and sampling diagnostics are summarized in Table~\ref{tab:si-reference-protocols}. Overlaps pair independently generated replicas at fixed disorder and temperature; statistical errors from correlated PT measurements must be assessed using independent chains or suitable blocks, rather than treating every recorded configuration as independent.

\begin{figure}[!htbp]
\centering
\includegraphics[width=\textwidth,height=0.58\textheight,keepaspectratio]{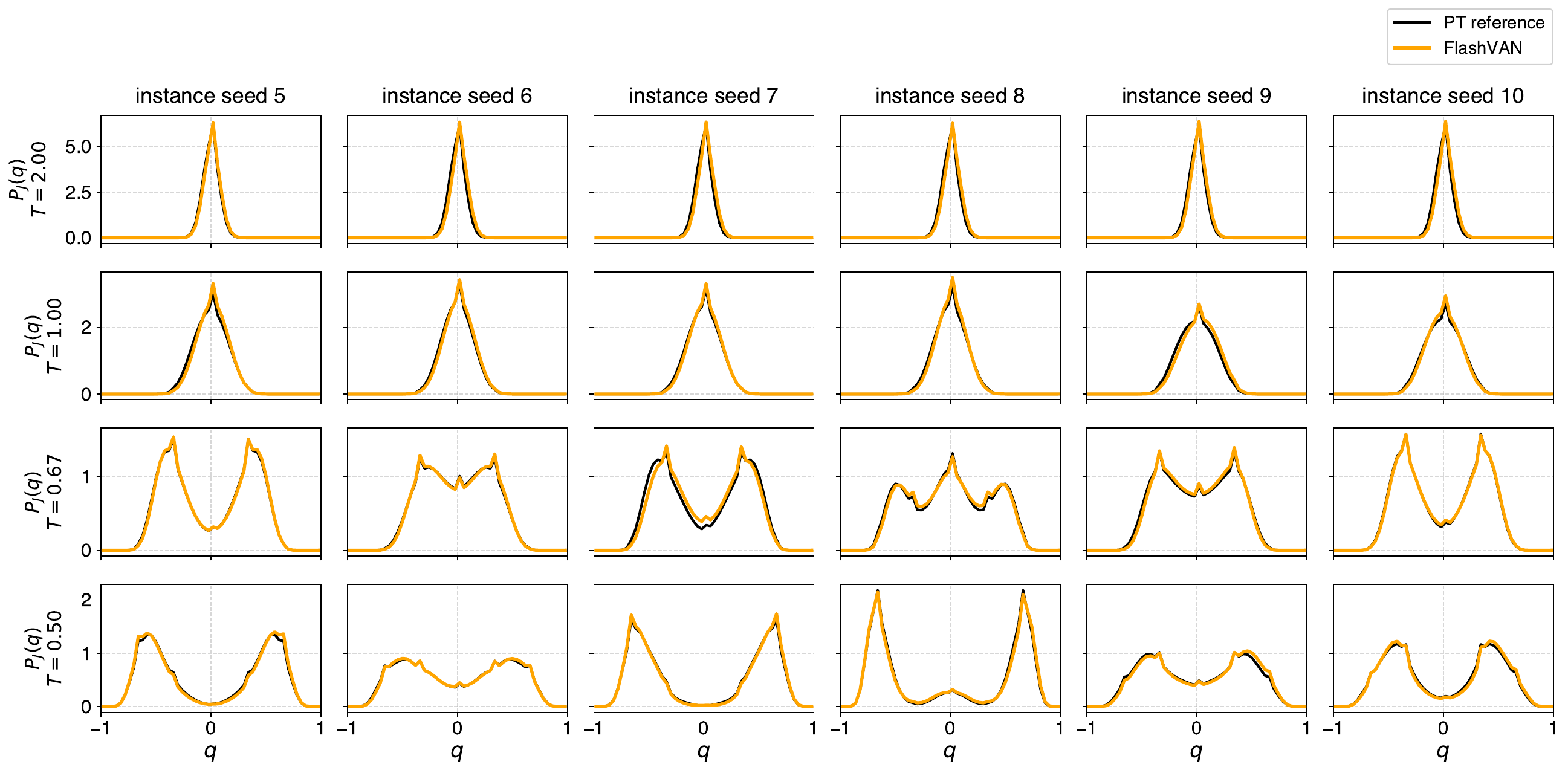}
\caption{\textbf{SK overlap distributions for six additional disorder instances.} Fixed-disorder $P_J(q)$ at $N=256$ and $T=2,1,2/3,1/2$, comparing FlashVAN with PT~\cite{responsePT1996} at the same couplings, size, and temperature. Seeds 5-10 supplement the four instances shown in the main text. Each model generates 20000 configurations directly, without LMC postprocessing or post-hoc spin-flip symmetrization.}
\label{fig:revised-sk-overlap}
\end{figure}

Figure~\ref{fig:revised-sk-overlap} extends the main-text SK overlap comparison to six additional disorder realizations, with coupling seeds 5-10. Together with the four main-text instances, these measurements cover ten realizations at the same system size and four temperatures. Matching the couplings and temperature between FlashVAN and PT tests the learned distribution for each individual instance.

The additional instances exhibit distinct low-temperature peak structures, which are closely reproduced by FlashVAN. The comparison extends the sampling assessment across disorder realizations rather than relying on the four main-text examples alone. The plotted configurations are sampled directly from the trained models, so the agreement reflects the generative distributions without subsequent Monte Carlo relaxation.

\FloatBarrier
\subsection{Matched SK optimization comparison with IAMP}\label{sec:si-iamp}
Table~\ref{tab:revised-iamp} reports an instance-matched energy comparison at $N=256$ between FlashVAN and IAMP~\cite{responseMontanariSK,responseIAMPImplementation2020}, an approximate-message-passing optimizer that carries an asymptotic optimization guarantee for the SK model. Using the same Gaussian couplings isolates differences in the low-energy configurations found by the two algorithms from fluctuations between disorder ensembles. At the largest IAMP budget, FlashVAN attains the lower discrete energy on 22 of the 30 instances and the remaining 8 are tied within the evaluation resolution; no instance has a lower IAMP energy beyond this threshold, and the mean signed difference is $+2.08\times10^{-3}\pm0.59\times10^{-3}$ per spin. This comparison concerns finite-size optimization accuracy only: it does not certify exact ground states, it does not address the asymptotic regime in which IAMP's guarantee applies, and computational costs are not compared. The Gaussian instances here are separate from the bimodal ensemble used for the main-text EO scaling comparison.

\begin{table}[!htbp]
\centering\small
\caption{Instance-matched ground-state-search energies per spin for FlashVAN and IAMP on 30 Gaussian SK instances at $N=256$. IAMP is shown at three budget settings: the four published discretizations $n_q=64,85,128,256$ with one initialization, the extended set adding $n_q=512,1024$ with one initialization, and the extended set with four initializations per discretization (24 runs per instance). The difference column uses the largest IAMP budget; positive values indicate a lower FlashVAN energy. Rows marked $^{\mathrm{t}}$ are ties, $|\Delta e|\le 10^{-6}$ per spin.}
\label{tab:revised-iamp}
\fitsitable{%
\begin{tabular}{cccccc}
\toprule
Instance & FlashVAN & \makecell{IAMP\\$n_q\le256$\\1 init} & \makecell{IAMP\\$n_q\le1024$\\1 init} & \makecell{IAMP\\$n_q\le1024$\\4 inits} & Difference\\
\midrule
1 & $-0.758672$ & $-0.757050$ & $-0.756249$ & $-0.758672$ & $+0.0000000^{\mathrm{t}}$\\
2 & $-0.741945$ & $-0.726642$ & $-0.729692$ & $-0.729692$ & $+0.0122525$\\
3 & $-0.754968$ & $-0.754353$ & $-0.754906$ & $-0.754906$ & $+0.0000623$\\
4 & $-0.747715$ & $-0.747175$ & $-0.738543$ & $-0.747715$ & $+0.0000000^{\mathrm{t}}$\\
5 & $-0.728552$ & $-0.718064$ & $-0.723057$ & $-0.723057$ & $+0.0054956$\\
6 & $-0.758062$ & $-0.752703$ & $-0.757020$ & $-0.758062$ & $-0.0000000^{\mathrm{t}}$\\
7 & $-0.737252$ & $-0.737182$ & $-0.732786$ & $-0.736917$ & $+0.0003352$\\
8 & $-0.741728$ & $-0.737984$ & $-0.731051$ & $-0.740788$ & $+0.0009397$\\
9 & $-0.741739$ & $-0.738409$ & $-0.738531$ & $-0.740721$ & $+0.0010183$\\
10 & $-0.755262$ & $-0.753590$ & $-0.754479$ & $-0.755262$ & $-0.0000000^{\mathrm{t}}$\\
11 & $-0.733374$ & $-0.731812$ & $-0.731050$ & $-0.732235$ & $+0.0011390$\\
12 & $-0.783808$ & $-0.783440$ & $-0.781030$ & $-0.782368$ & $+0.0014393$\\
13 & $-0.747780$ & $-0.731190$ & $-0.736974$ & $-0.737063$ & $+0.0107172$\\
14 & $-0.742651$ & $-0.740658$ & $-0.737608$ & $-0.742542$ & $+0.0001094$\\
15 & $-0.750217$ & $-0.747767$ & $-0.746876$ & $-0.748146$ & $+0.0020704$\\
16 & $-0.737511$ & $-0.728158$ & $-0.734879$ & $-0.735601$ & $+0.0019100$\\
17 & $-0.733482$ & $-0.728021$ & $-0.728975$ & $-0.730650$ & $+0.0028320$\\
18 & $-0.753524$ & $-0.740627$ & $-0.752950$ & $-0.753342$ & $+0.0001820$\\
19 & $-0.755445$ & $-0.751888$ & $-0.752308$ & $-0.755445$ & $-0.0000000^{\mathrm{t}}$\\
20 & $-0.747582$ & $-0.743430$ & $-0.745448$ & $-0.747128$ & $+0.0004544$\\
21 & $-0.745809$ & $-0.738688$ & $-0.743428$ & $-0.743428$ & $+0.0023814$\\
22 & $-0.727286$ & $-0.722535$ & $-0.722725$ & $-0.727148$ & $+0.0001375$\\
23 & $-0.757434$ & $-0.752646$ & $-0.755494$ & $-0.757434$ & $+0.0000000^{\mathrm{t}}$\\
24 & $-0.751379$ & $-0.748425$ & $-0.749532$ & $-0.749532$ & $+0.0018463$\\
25 & $-0.766491$ & $-0.764871$ & $-0.766491$ & $-0.766491$ & $+0.0000000^{\mathrm{t}}$\\
26 & $-0.755648$ & $-0.749292$ & $-0.754417$ & $-0.754417$ & $+0.0012311$\\
27 & $-0.746728$ & $-0.744601$ & $-0.744653$ & $-0.746728$ & $-0.0000000^{\mathrm{t}}$\\
28 & $-0.736631$ & $-0.725885$ & $-0.733144$ & $-0.734910$ & $+0.0017209$\\
29 & $-0.736047$ & $-0.722875$ & $-0.721337$ & $-0.727512$ & $+0.0085348$\\
30 & $-0.738227$ & $-0.733317$ & $-0.732744$ & $-0.732744$ & $+0.0054831$\\
\midrule
Mean & $-0.747098$ & $-0.741776$ & $-0.742946$ & $-0.745022$ & $+0.0020764\pm0.0005911$\\
\bottomrule
\end{tabular}}
\end{table}
The comparison uses identical zero-diagonal Gaussian matrices and Hamiltonian normalization, with instances drawn from the reference implementation's Gaussian matrix distribution under a seeded generator for reproducibility. IAMP~\cite{responseMontanariSK,responseIAMPImplementation2020} uses the authors' precomputed zero-temperature order parameter (250 RSB discretization levels) and their default Parisi-PDE grid ($x_{\max}=10$, $n_x=500$); it is evaluated at discretizations $n_q=64,85,128,256,512,1024$, i.e. step sizes $\delta=1/n_q$ corresponding to relative scales $N\delta=4,3,2,1,\tfrac12,\tfrac14$ at $N=256$. The table lists the best discrete energy at three budgets: the four published discretizations with one initialization, the extended set with one initialization, and the extended set with four independent random initializations per discretization. The budget settings use independent random initializations, so their outcomes need not improve monotonically from one column to the next. IAMP's terminal vector is converted to spins by its componentwise sign. Saved spin configurations from both methods are reevaluated in double precision. The difference is $e_{\mathrm{IAMP}}-e_{\mathrm{FlashVAN}}$, and values with $|\Delta e|\le 10^{-6}$ per spin are classified as ties, a threshold set by the single-precision resolution of the training-side energy evaluation. With a single initialization per discretization, FlashVAN attains the lower energy on all 30 instances at the published four-grid budget (minimum difference $7\times10^{-5}$ per spin) and on 29 of 30 instances at the extended budget. The eight ties at the largest budget reflect IAMP's run-to-run spread at fixed instance and grid (median $4.4\times10^{-3}$ per spin across the four initializations), which is comparable to the mean inter-method difference.

\FloatBarrier
\subsection{EA reference data and overlap evaluation}
\label{sec:si-ea-references}
For the 2D EA overlap comparison, the four fixed disorder instances and PT overlap reference data are from Del Bono et al.~\cite{del2025nearest}. The Kac-Ward reference data were computed by our collaborators using the exact autoregressive sampling code of Liu et al.~\cite{responseKacWardSampler2026}. The couplings and temperature are retained in the normalization of the supplied instances. These instance-specific tests are separate from the independently generated Gaussian ensembles used for energy benchmarks. McGroundState~\cite{CJMM22} supplies the exact 2D ground-state references. The main-text energy residual uses the final sample-mean energy density and is aggregated geometrically over disorder instances.

The 3D EA $L=10$ overlap comparison uses the fixed instance and PT/PA reference distributions from Del Bono et al.~\cite{del2026demonstrating}. FlashVAN configurations are sampled directly after distillation. For consistency with this reference, the plotted 3D $L=10$ overlap histograms are symmetrized under global spin reversal. The SK distributions are plotted without such symmetrization. Table~\ref{tab:si-reference-protocols} records the reference-sampling and histogram settings.

For 3D EA energy scaling, the PT benchmark values are taken from Rom{\'a} et al.~\cite{ROMA20092821} and the tabulation of Wang et al.~\cite{responseWangGroundStates2015}. The finite-size lines in the main-text energy panels are literature references~\cite{boettcher2023deep}, rather than fits to FlashVAN data. The SK thermodynamic-limit RSB reference does not supply an exact prediction for finite-dimensional 3D EA.

\begin{table}[!htbp]
\centering\small
\setlength{\tabcolsep}{4pt}
\renewcommand{\arraystretch}{1.15}
\caption{Reference data and overlap evaluation settings.}
\label{tab:si-reference-protocols}
\fitsitable{%
\begin{tabular}{lccc}
\toprule
Parameter & SK overlaps & 2D EA overlaps & 3D EA overlaps\\
\midrule
Reference source & PT~\cite{responsePT1996} & \makecell{PT data: Ref.~\cite{del2025nearest}\\Kac-Ward: Ref.~\cite{responseKacWardSampler2026}} & \makecell{PT/PA from\\Ref.~\cite{del2026demonstrating}}\\
Couplings / size & Matched; $N=256$ & Reference instances; $L=16$ & Reference instance; $L=10$\\
Temperature & $2,1,2/3,1/2$ & $0.3125$ & $1.92,0.77,0.29,0.10$\\
PT independent chains & 32 & See reference & See reference\\
PT replica count / ladder & 27; geometric $\beta\in[0.2,2]$ & See reference & See reference\\
PT equilibration sweeps & 20000 & See reference & See reference\\
PT measurement sweeps / thinning & 60000 / 5 & See reference & See reference\\
PT equilibration diagnostics & \makecell{Round trips ($\ge27$ per chain);\\energy half-drift ($\le0.35$)} & See reference & See reference\\
Direct FlashVAN configurations & 20000 & 200000 & 20000\\
Global spin-flip symmetrization & No & No & Yes\\
\bottomrule
\end{tabular}}
\par\smallskip
\raggedright\footnotesize The 2D EA PT data are supplied from Ref.~\cite{del2025nearest}; the Kac-Ward data were computed by collaborators using the code of Ref.~\cite{responseKacWardSampler2026}. The 3D PT/PA distributions are taken from Ref.~\cite{del2026demonstrating}.
\end{table}

\subsection{Small-overlap probabilities in 3D EA}\label{sec:si-small-overlap}
The small-overlap probability is computed for each disorder instance from 10000 disjoint pairs of 20000 direct model configurations, followed by an average over instances. The $L=14$ study uses 60 instances at each of the three lower temperatures and ten at each of the four higher temperatures. The $L=16$ extension uses ten per temperature. Bootstrap resampling acts on disorder instances; it does not resample the 10000 pairs as if they were independent disorder samples.

\begin{figure}[!htbp]
\centering
\includegraphics[width=0.5\textwidth,height=0.58\textheight,keepaspectratio]{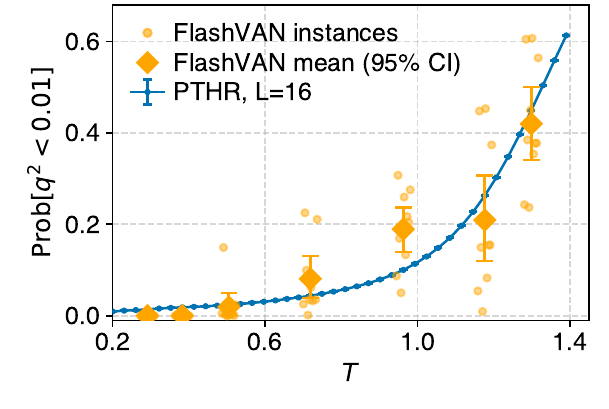}
\caption{\textbf{Small-overlap probabilities in 3D EA at $L=16$.} FlashVAN is compared with the independent PTHR reference~\cite{chilin2026true,responseChilinPTHR2026} at seven temperatures, using ten disorder instances per temperature. Pale circles denote individual-instance probabilities, and diamonds denote their mean with 95\% disorder-bootstrap confidence intervals. The intervals are percentile intervals from 10000 resamples of disorder instances. Each trained model supplies 20000 direct configurations and 10000 disjoint replica-pair overlaps, without LMC postprocessing. The PTHR curve and its supplied errors use a different disorder ensemble.}
\label{fig:revised-l16-overlap}
\end{figure}

The main-text $L=14$ error bars are bootstrap standard errors. Figure~\ref{fig:revised-l16-overlap} instead displays 95\% percentile confidence intervals from 10000 disorder-bootstrap resamples. The PTHR references~\cite{chilin2026true,responseChilinPTHR2026} use independent disorder ensembles and the nearest published temperature points. Their quoted uncertainties are not included in the FlashVAN bootstrap intervals.

To examine the size dependence of the small-overlap benchmark, we extend the $L=14$ comparison to $L=16$ in Fig.~\ref{fig:revised-l16-overlap}. The seven-temperature study uses ten disorder instances per temperature and the same population-assisted training components as the main-text small-overlap experiment. Showing individual-instance probabilities alongside their mean makes the disorder fluctuations visible.

The small-overlap weight increases with temperature, following the overall trend of the PTHR (Parallel Tempering enhanced with Houdayer moves and with an entropic reservoir) reference~\cite{chilin2026true,responseChilinPTHR2026}. Quantitative agreement remains temperature dependent: the lowest-temperature model probabilities remain below the reference, while some intermediate-temperature means are higher. This comparison characterizes both the feasibility of a $4096$-spin generative sampler and the remaining challenge of reproducing low-overlap weights as system size increases.

\clearpage
\section{Computational performance and neural baselines}
\label{sec:hyparameters_details}
\subsection{Hardware, precision, and implementation benchmarks}
\label{sec:hardware}
\begin{figure}[!htbp]
	\begin{center}
		\includegraphics[width=\columnwidth,height=0.58\textheight,keepaspectratio]{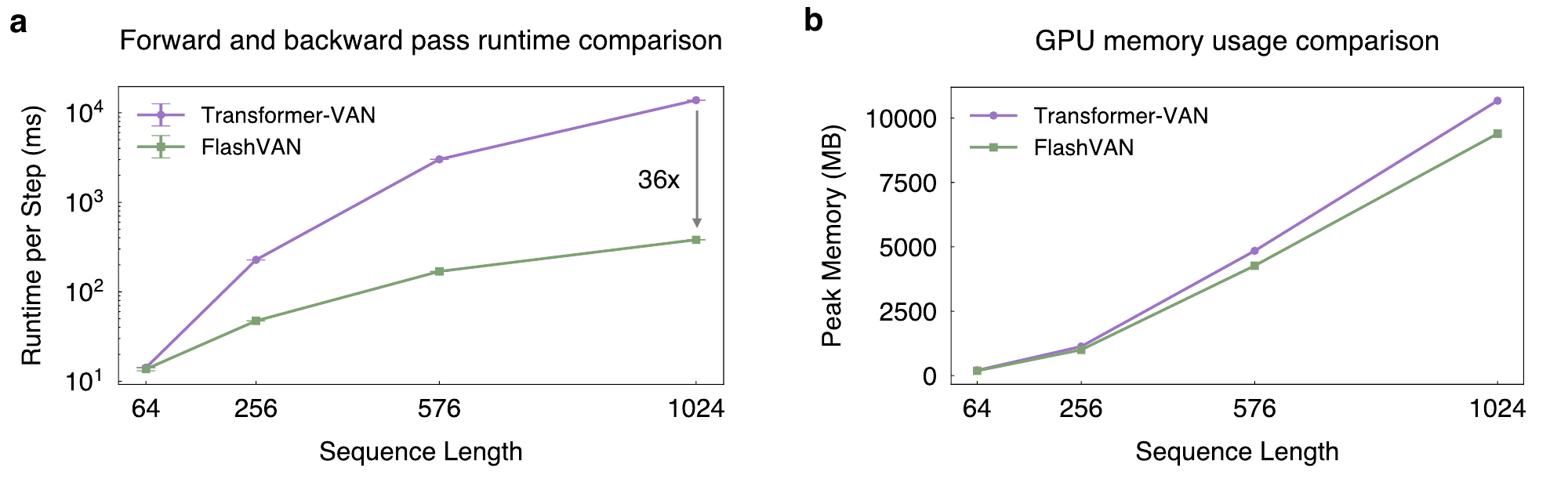}
	\end{center}
	\caption{\textbf{Performance comparison between the transformer-based VAN (purple) and FlashVAN (green).} (a) Mean per-step runtime for the combined forward and backward pass, averaged over 500 steps. (b) Peak GPU memory usage during training.}
	\label{fig2C}
\end{figure}

The controlled implementation benchmarks in Fig.~\ref{fig2C} were conducted on a single NVIDIA H100 GPU (80 GB); the population-assisted sampling, optimization, and new timing measurements use an NVIDIA RTX~5090 GPU. The energy-scaling datasets in main-text Figs.~3c and~4c retain their separately specified annealing configurations. Timings obtained on different devices are not combined into a single speedup estimate. Attention kernels use FP16 inputs, whereas embeddings, feed-forward operations, and energy evaluation use FP32. Saved configurations in the IAMP comparison are reevaluated in double precision.

Figure~\ref{fig2C} compares a Transformer VAN using naive PyTorch attention with FlashVAN. Its panels report forward/backward runtime and training memory; these module-level measurements are distinct from the full training-plus-sampling search costs in Section~\ref{sec:si-ga-costs}.

For full causal attention over $n$ tokens, a single attention evaluation requires $\mathcal O(n^2d)$ work. Recomputing the entire prefix at every generation step leads to $\mathcal O(n^3d)$ cumulative work. KV caching reuses preceding keys and values and reduces cumulative attention work to $\mathcal O(n^2d)$. FlashAttention~\cite{Dao_2024} reduces memory traffic and intermediate storage without changing the asymptotic full-attention complexity. The sparse-attention measurements in Fig.~\ref{fig1A} separately examine the effect of restricting the mask.

\FloatBarrier
\subsection{Matched ground-state-search costs}\label{sec:si-ga-costs}
Table~\ref{tab:revised-ga-times} reports the individual-instance times underlying the main-text optimization-cost comparison. The median times are comparable, with FlashVAN faster on five instances and GA faster on five. The variation across instances shows why the median comparison is accompanied by the full set of matched results.

\begin{table}[!htbp]
\centering\small
\caption{Ground-state-search times for ten matched $L=14$ 3D EA instances on an RTX~5090.}
\label{tab:revised-ga-times}
\begin{tabular}{ccc}
\toprule
Instance seed & FlashVAN $t_{0.9}$ (s) & GA rerun time $t_{0.9}$ (s)\\
\midrule
502033 & 419 & 717\\
9569658 & 411 & 656\\
7711527 & 480 & 629\\
14949383 & 547 & 658\\
3210623 & 684 & 730\\
9551411 & 819 & 773\\
16095495 & 820 & 733\\
5105474 & 819 & 658\\
4063355 & 2683 & 815\\
11086240 & 3375 & 1048\\
\midrule
Median & 752 & 724\\
\bottomrule
\end{tabular}
\end{table}
For each tested budget, success means that the best observed energy density agrees with the reference to five decimal places. The attainment time $t_{0.9}$ is the mean runtime at the first tested budget reaching at least 90\% success. FlashVAN timing includes the applicable training stages; initial high-temperature population preparation is excluded following the reference convention~\cite{del2026demonstrating}. GA entries are the mean runtimes of three RTX~5090 reruns at the original settings first attaining 90\% success. In the main-text curves, the published V100 runtimes are divided by the measured hardware ratio 2.7905; GA and PA markers use directly timed RTX~5090 reruns.

\subsubsection{Transferring population-assisted training to MADE}
\label{sec:si-made-transfer}
\begin{figure}[!htbp]
\centering
\includegraphics[width=0.5\columnwidth]{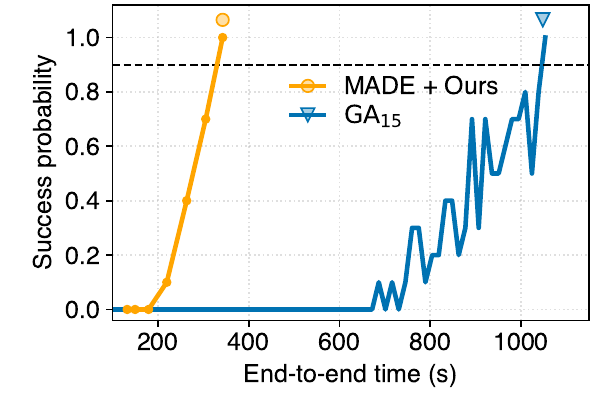}
\caption{\textbf{Population-assisted training with MADE on the $L=14$ instance 11086240.} Each MADE budget is evaluated with ten runs. The first budget attaining at least 90\% success gives 10/10 successes at 700 annealing updates, with mean recorded runtime 342.6 s. The GA curve uses calibrated published V100 times; the triangle marks the directly measured RTX~5090 reference of 1048 s~\cite{del2026demonstrating}. The timing scope is specified in the text.}
\label{fig:si-made-transfer}
\end{figure}

To examine whether the training strategy can be used with a simpler autoregressive architecture, we replace the Transformer in the $L=14$ ground-state-search protocol by the masked-linear MADE architecture used in GA~\cite{del2026demonstrating,germain2015made}. The test uses seed 11086240, which has the longest GA attainment time among the ten benchmark instances. Figure~\ref{fig:si-made-transfer} and Table~\ref{tab:si-made-transfer} report this instance-specific comparison.

The model has a strictly lower-triangular masked matrix $W$ without bias, with spin conditionals
\begin{equation}
q_\theta(\sigma_i=+1\mid\sigma_{<i})=
\operatorname{sigmoid}\!\left(2\sum_{j<i}W_{ij}\sigma_j\right).
\end{equation}
Spins are generated individually in lattice raster order. The model uses no positional embedding or attention. A reversible conversion to $2\times2\times2$ patch codes supplies the interface to the existing pool implementation, without changing the spin-wise autoregressive probabilities.

We use the same population-assisted training strategy and evaluation protocol as the FlashVAN ground-state-search benchmark in main-text Fig.~4d (Table~\ref{tab:si-3dea-protocols}), using MADE with logarithmic annealing.

We test annealing budgets of 100, 200, 300, 400, 500, 600, and 700 updates, with ten runs at each budget. A run succeeds when the best recorded energy density rounds to the reference $-1.68613$ at five decimal places. The first tested budget reaching at least 90\% success is 700 updates, with ten successes out of ten and a mean recorded runtime of 342.6 s. The directly timed GA reference is 1048 s, and the Transformer-based FlashVAN reference is 3375 s for this instance.

The timing follows the GA benchmark convention: the recorded MADE runtime includes pretraining, annealing, and refinement, excluding initial population preparation and final autoregressive generation. The plotted GA curve uses published V100 times divided by 2.7905, while its marker shows the directly measured RTX~5090 attainment time. This result shows that the population-assisted training strategy is effective with a simpler autoregressive architecture on this instance. The main FlashVAN benchmarks retain the Transformer to address finite-temperature overlap statistics and larger-system probability modeling alongside optimization.

\begin{table}[!htbp]
\centering\small
\caption{Recorded optimization times and MADE budget scan for seed 11086240.}
\label{tab:si-made-transfer}
\begin{tabular}{lcc}
\toprule
Method & Attainment statistic & Recorded time (s)\\
\midrule
MADE + population-assisted training & $t_{0.9}$ & 342.6\\
FlashVAN & $t_{0.9}$ & 3375\\
GA, directly timed RTX~5090 & $t_{0.9}$ & 1048\\
\bottomrule
\end{tabular}
\par\medskip
\begin{tabular}{ccc}
\toprule
MADE annealing updates & Successful runs & Mean recorded time (s)\\
\midrule
100 & 0/10 & 132.3\\
200 & 0/10 & 149.7\\
300 & 0/10 & 179.4\\
400 & 1/10 & 219.5\\
500 & 4/10 & 263.4\\
600 & 7/10 & 305.3\\
700 & 10/10 & 342.6\\
\bottomrule
\end{tabular}
\end{table}

\subsection{Training times for overlap measurements}
Table~\ref{tab:revised_overlap_training_time} reports the mean training time per disorder instance for each model, system size, and target temperature. The number of disorder realizations contributing to each mean is listed in the No. of instances column; each model is trained for a fixed realization and temperature. The total includes pretraining and all subsequent training stages used by that protocol, including Monte Carlo refinement of the training population; generation of the configurations used for final observable measurements is considered separately.

These times characterize the initial cost of constructing the sampler. Once training is complete, the same model can generate additional configurations for repeated measurements at the specified disorder and temperature. The separation between training and subsequent sampling therefore helps assess the computational cost for different measurement workloads.

\begin{table}[!htbp]
\centering
\small
\setlength{\tabcolsep}{4pt}
\renewcommand{\arraystretch}{1.15}
\caption{Mean total training time per disorder instance for the overlap benchmarks, averaged over the indicated numbers of disorder realizations.}
\label{tab:revised_overlap_training_time}
\begin{tabular}{ccccc}
\toprule
Model & System size & Target $T$ & No. of instances & Mean training time (min)\\
\midrule
\multirow[c]{4}{*}{SK} & \multirow[c]{4}{*}{$N=256$} & $2$ & $10$ & $31.5$\\
 &  & $1$ & $10$ & $33.4$\\
 &  & $2/3$ & $10$ & $33.5$\\
 &  & $1/2$ & $10$ & $36.0$\\
\midrule
\multirow[c]{1}{*}{2D EA} & \multirow[c]{1}{*}{$L=16$} & $0.3125$ & $4$ & $2.8$\\
\midrule
\multirow[c]{18}{*}{3D EA} & \multirow[c]{4}{*}{$L=10$} & $1.92$ & $1$ & $12.2$\\
 &  & $0.77$ & $1$ & $25.0$\\
 &  & $0.29$ & $1$ & $20.4$\\
 &  & $0.10$ & $1$ & $22.8$\\
\cmidrule(l){2-5}
 & \multirow[c]{7}{*}{$L=14$} & $1.2983$ & $10$ & $35.6$\\
 &  & $1.1763$ & $10$ & $41.7$\\
 &  & $0.9627$ & $10$ & $61.8$\\
 &  & $0.7186$ & $10$ & $55.5$\\
 &  & $0.5051$ & $60$ & $44.9$\\
 &  & $0.3831$ & $60$ & $44.3$\\
 &  & $0.2915$ & $60$ & $45.8$\\
\cmidrule(l){2-5}
 & \multirow[c]{7}{*}{$L=16$} & $1.2983$ & $10$ & $56.4$\\
 &  & $1.1763$ & $10$ & $84.6$\\
 &  & $0.9627$ & $10$ & $99.8$\\
 &  & $0.7186$ & $10$ & $95.2$\\
 &  & $0.5051$ & $10$ & $92.0$\\
 &  & $0.3831$ & $10$ & $71.2$\\
 &  & $0.2915$ & $10$ & $70.6$\\
\bottomrule
\end{tabular}
\end{table}

\FloatBarrier
\subsection{Comparison with MADE}\label{sec:compare_with_made}

\subsubsection{Model configurations}

MADE~\cite{germain2015made} is a feed-forward autoregressive network whose masked connections enforce the chosen generation order. Table~\ref{tab:made-settings} lists the MADE architectures and parameter counts used in this controlled comparison; the plotted FlashVAN configurations vary the embedding dimension and number of heads. Patching groups nearby spins into joint categorical tokens for both networks. These experiments compare neural representations and per-step computation; the matched comparisons with conventional solvers are reported separately.

\begin{table}[!htbp]
\centering
\caption{Trainable parameter statistics of MADE.}
\label{tab:made-settings}
\begin{tabular}{
  l
  c
  c
  c
  c
}
\toprule
System size & Net depth & Net width & {Parameters (w/ patch)} & {Parameters (w/o patch)}  \\
\midrule
\multirow{6}{*}{L = 8}
  & 1 & 1  & 4\,096      & 4\,096 \\
  & 2 & 1  & 4\,368    & 8\,256    \\
  & 2 & 4  & 17\,472   & 33\,024    \\
  & 2 & 8  & 34\,944   & 66\,048   \\
  & 2 & 16 & 69\,888   & 132\,096  \\
  & 3 & 4  & 25\,728   & 164\,352  \\
\midrule
\multirow{5}{*}{L = 32}
  & 1 & 1  & 1\,048\,576  & 1\,048\,576 \\
  & 2 & 4  & 4\,457\,472  & 8\,392\,704 \\
  & 2 & 16 & 17\,829\,888 & 33\,570\,816 \\
  & 2 & 32 & 35\,659\,776 & \multicolumn{1}{c}{-} \\
  & 3 & 4  & 6\,555\,648  & \multicolumn{1}{c}{-} \\
\bottomrule
\end{tabular}
\end{table}

\subsubsection{Accuracy and per-step runtime}

\begin{figure}[!htbp]
	\begin{center}
		\includegraphics[width=\columnwidth,height=0.58\textheight,keepaspectratio]{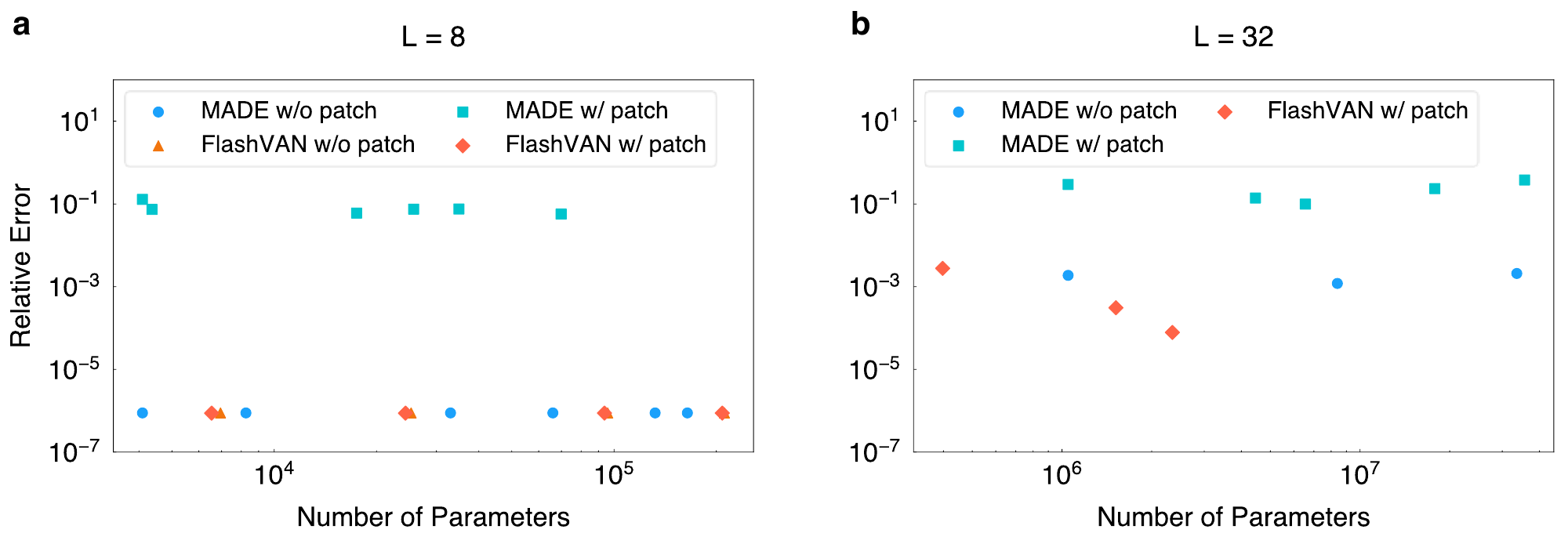}
	\end{center}
	\caption{\textbf{Benchmarking the accuracy of MADE and FlashVAN on the 2D EA model.} Panels (a) and (b) show the residual energy per site as a function of the number of trainable parameters. Each data point is averaged over five disorder instances. (a) System size $L = 8$. (b) System size $L = 32$.}
	\label{fig4B_1}
\end{figure}

For the 2D EA ground-state estimation task, we evaluate both convergence accuracy and computational cost for MADE and FlashVAN. Supplementary Fig.~\ref{fig4B_1} presents the absolute energy residual per spin relative to the exact energies, averaged over five disorder instances, as a function of the number of trainable parameters. At $L=8$, FlashVAN attains an small energy residual ($\approx 10^{-6}$), retaining high precision even under patching configurations. In contrast, MADE achieves comparable accuracy only in its unpatched form; once patched, its error deteriorates to $10^{-1}$, indicating that input compression degrades its ability to preserve key information. This limitation becomes more pronounced at $L=32$: despite scaling to tens of millions of parameters, MADE fails to reach low-error solutions and is consistently outperformed by FlashVAN in both patched and unpatched regimes.

\begin{figure}[!htbp]
	\begin{center}
		\includegraphics[width=\columnwidth,height=0.58\textheight,keepaspectratio]{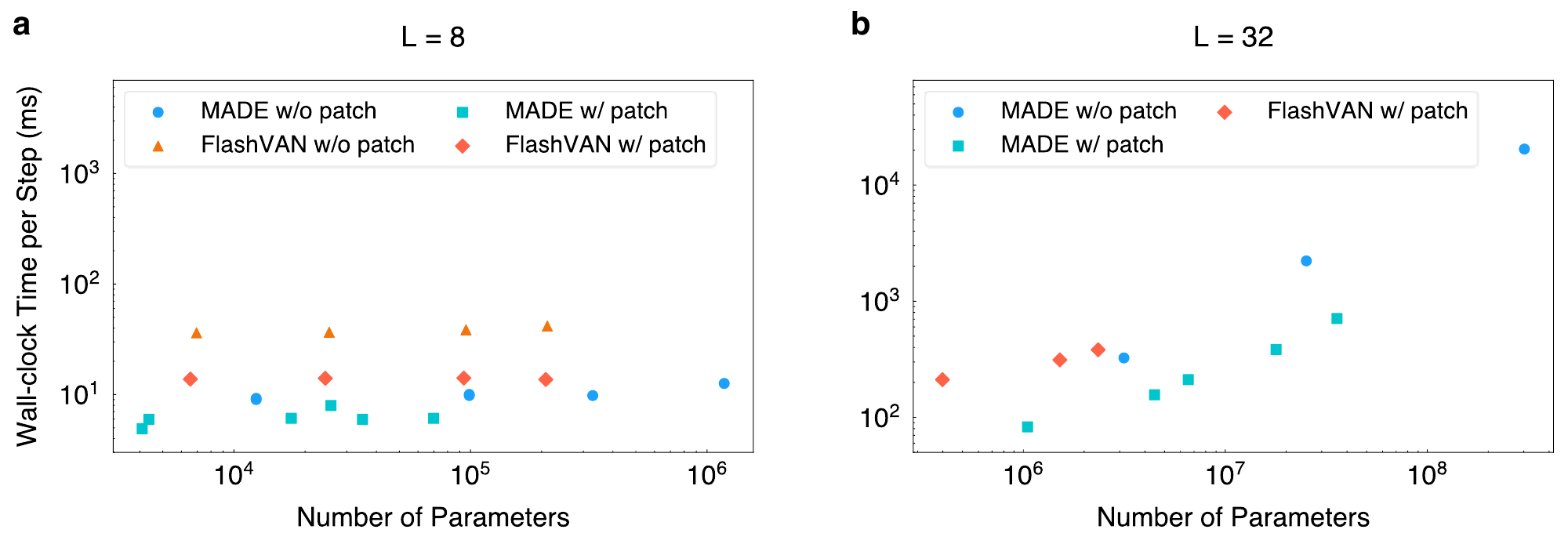}
	\end{center}
	\caption{\textbf{Comparison of wall-clock training efficiency between MADE and FlashVAN.} Panels (a) and (b) show the wall-clock time per training step as a function of the number of trainable parameters for system sizes $L = 8$ and $L=32$, respectively. Each data point is averaged over 1,000 training steps.}
	\label{fig4B_2}
\end{figure}

Figure~\ref{fig4B_2} reports per-step wall-clock times averaged over 1000 steps. FlashVAN has slightly higher per-step overhead than MADE at comparable parameter scales, while the accuracy comparison in Fig.~\ref{fig4B_1} shows the benefit of the attention-based representation in the tested configurations. Per-step timings alone do not establish a lower total cost at matched physical accuracy.

\clearpage
\bibliography{SI-FlashVAN}